\documentclass[aps,pra, twocolumn, showpacs,floatfix,nofootinbib,superscriptaddress]{revtex4-2}
\usepackage{amsmath,amssymb}
\usepackage{graphicx}
\usepackage{appendix}
\usepackage{subcaption}
\usepackage[usenames,dvipsnames]{color}
\usepackage[colorlinks=true, citecolor=blue, linkcolor=blue, urlcolor=blue]{hyperref}
\usepackage{hyphenat}
\usepackage{txfonts}
\usepackage{dsfont}
\usepackage{enumitem}
\usepackage[utf8]{inputenc}

\usepackage[nottoc,numbib]{tocbibind}

\newcommand{\ep}{{\epsilon}}
\newcommand{\Ea}{{E_\alpha}}
\newcommand{\PEa}{{ \hat{\mathcal{P}}_{\Ea} }}
\newcommand{\omL}{{\omega_L}}
\newcommand{\omc}{{\omega_c}}

\begin{document}

\title{Excitonic enhancement of cavity-mediated interactions in a two-band Hubbard model}

\author{Xiao Wang}
\affiliation 
{Department of Physics, University of Oxford, UK}
\affiliation 
{Max Planck Institute for the Structure and Dynamics of Matter, Luruper Chaussee 149, 22761 Hamburg, Germany}

\author{Dieter Jaksch}
\affiliation{University of Hamburg, Luruper Chaussee 149, Hamburg, Germany}
\affiliation{The Hamburg Centre for Ultrafast Imaging, Hamburg, Germany}
\affiliation 
{Department of Physics, University of Oxford, UK}

\author{Frank Schlawin}
\affiliation 
{Max Planck Institute for the Structure and Dynamics of Matter, Luruper Chaussee 149, 22761 Hamburg, Germany}
\affiliation{University of Hamburg, Luruper Chaussee 149, Hamburg, Germany}
\affiliation{The Hamburg Centre for Ultrafast Imaging, Hamburg, Germany}

\date{\today}

\begin{abstract}

We study cavity-mediated interactions that are generated in a two-dimensional two-band Hubbard model coupled to an optical cavity, when it is driven in-gap by a strong laser. 
Starting from a Floquet description of the driven system, 
we derive effective low-energy Hamiltonians by projecting out the high-energy degrees of freedom and treating intrinsic interactions on a mean field level. We then investigate how the emergence of high-energy Frenkel excitons from the electronic interband coupling, which form near the upper electronic band, affects the interactions as well as the laser-induced Floquet renormalization of the electronic band structure. 
Cavity-mediated interactions are enhanced strongly when the light couples to an excitonic transition. 
Additionally, the interaction as well as the Floquet renormalization are strongly broadened in reciprocal space, which could further boost the impact of cavity-mediated interactions on the driven-dissipative steady state. 

\end{abstract}

\maketitle

\section{Introduction}

The laser control of low-energy degrees of freedom in complex quantum many-body systems has emerged as one of the most active fields of condensed matter and many-body physics. It is referred to as Floquet engineering of quantum materials~\cite{oka2019floquet,Eckardt2017}, if the control is based on the coherent manipulation of the system's degrees of freedom, or optical switching~\cite{RevModPhys.93.041002}, if it relies on the deposition of energy and e.g. the melting of competing order in a transient nonthermal state.
 Laser excitation can stabilize coherent phases such as superconductivity above their equilibrium critical temperature~\cite{Fausti2011, Mitrano2016, Budden2021}, or transiently change the crystal structure to unlock new ground states~\cite{Foerst2011, Disa2021}.
Moreover, new phenomena, which are sometimes not present in the equilibrium material, such as 
Floquet-induced topology~\cite{rudner2020band},
or transient superconductivity~\cite{Buzzi2020,Tindall2020, Tindall2021,Tindall2021liebstheorem}, have been reported. 

Recently, cavity control of quantum materials~\cite{schlawin2022cavity,bloch2022strongly, garcia2021manipulating, Mivehvar2021, Ruggenthaler2018}, where the quantum fluctuations of light in an undriven cavity act as a tailored environment of the material, is also receiving extensive interest.
First experiments report, inter alia, the polaritonic manipulation of chemical reaction rates~\cite{garcia2021manipulating}, cavity-modified carrier mobility in organic semiconductors~\cite{orgiu2015conductivity} and in Landau levels~\cite{Paravicini-Bagliani2019}, the change of superconducting critical temperatures~\cite{ThomasAnoop2019ESuS} and a metal-insulator transition~\cite{jarc2022cavity}, and the breakdown of topological protection of edge states under strong light-matter coupling~\cite{Appugliese2022}.
Theoretically proposed effects include cavity-mediated long-range interactions~\cite{PhysRevLett.122.133602, andolina2022, Chakraborty2021, Ciuti2021, Schaefer2021}, the renormalisation of electronic bandwidths~\cite{PhysRevB.105.165121,sentef2020quantum} and magnetic interaction strengths~\cite{Kiffner2019b, Curtis2022}, the shift of 
phase transitions~\cite{PhysRevLett.122.167002, doi:10.1126/sciadv.aau6969, Li2020, PhysRevX.10.041027, Latini2021, Passetti2023},
or the opening of topological gaps~\cite{Xiao2019, Hubener2021,dmytruk2022controlling} when the cavity is coupled off-resonantly to a quantum material. 
Resonant coupling may generate exotic superconductor-polaritons~\cite{PhysRevB.99.020504} or Mott polaritons~\cite{Kiffner2019}, and enable the control of exciton properties~\cite{Latini2019}. 
Transport properties of excitonic~\cite{schachenmayer2015cavity} or electronic systems~\cite{Bartolo2018, doi:10.1021/acs.jpcb.0c03227, Rokaj2019, Rokaj2022,Arwas2023} are predicted to be strongly influenced. 
In an ultrastrong coupling regime, where nonperturbative light-matter coupling may give rise to a superradiant phase transitions~\cite{Mazza2019, Andolina2019, Andolina2020, RomanRoche2021, SaezBlazquez2023}, exotic many-body phases of matter are predicted to emerge~\cite{Guerci2020, Rao2023, Rokaj2023, mercurio2023photon}.

When a hybrid cavity-many-body system is driven by lasers, more exotic Floquet engineering effects can emerge, which can be realised by neither bare-laser control nor bare-cavity control. First 
pioneering applications include the generation of supersolid phases of matter through the use of cavity-mediated long-range interactions in cold atoms~\cite{LandigRenate2016Qpfc} or room-temperature exciton-polariton condensation in 2d semiconductor heterostructures~\cite{liu2015strong,schneider2016exciton}. 
Recent theory work further proposes important new directions for material control, which include photo-assisted tunable electron pairing in the Cooper channel~\cite{GaoHongmin2020Pepi}, laser-assisted cavity-mediated topological superfluidity~\cite{SchlawinFrank2019CUPi}, cavity-induced quantum spin liquids~\cite{chiocchetta2021cavity}, and a largely unexplored crossover between the quantum and a classical Floquet regime~\cite{sentef2020quantum}. 

In this paper we show how electronic interactions can strongly enhance Floquet engineering effects and cavity-mediated interactions between electrons in the valence band of a weakly hole-doped semiconductor.  
We focus on the dispersive light-matter coupling regime, where no real excitations are created by the driving. 
Yet even in this case, we find that when the detuning between light and excitonic resonances becomes smaller than the exciton binding energy, many-body effects become important and the treatment of the excitons as effective two-level emitters breaks down. This finding could have important consequences for the optical control of quantum materials.
We have checked that our results encompass the results one obtains from a generalized RPA (GRPA) calculation of near-resonant light-matter processes. 
In addition, our approach allows us to describe strong-field effects such as the Bloch-Siegert shift, which cannot be captured in the GRPA method, and it can be extended to include a self-consistent treatment of the low-energy degrees of freedom, when the adiabaticity condition is violated. 

In the itinerant electron system considered here, the cavity-mediated interaction takes the form of a forward-scattering density-density interaction~\cite{chiocchetta2021cavity}. 
It can be enhanced by a small laser-cavity detuning and a large laser driving strength, and thereby can become much stronger than direct cavity-induced changes to the ground state, 
which require ultrastrong coupling to become relevant~\cite{frisk2019ultrastrong}. 
This scheme is thus very similar to established scheme in atomic cavity  QED~\cite{Mivehvar2021}.  
This regime separates our work from the highly active research on exciton-polaritons~\cite{doi:10.1126/science.aac9439,RevModPhys.82.1489,sanvitto2016road,dirnberger2022spin}, which is mainly concerned with a resonant coupling regime, and from Floquet engineering of excitons~\cite{Iorsh2022, Kobayashi2023, Conway2023}, which is concerned with changes of the excitonic states rather than the low-energy degrees of freedom. 
To describe the impact of cavity-mediated interactions on the electronic many-body state, they have to be treated on equal footing with other laser-induced Floquet engineering effects, namely the Stark and Bloch-Siegert shifts~\cite{Sie2018}, which are unavoidable. In addition, all these effects are screened by electronic interactions, which will further affect their magnitude and range. 
Here we explore this interplay for the specific case of screening by local interband interactions in a two-band semiconductor model, which give rise to Frenkel excitons.  
We extend a Floquet approach developed by Vogl et al.~\cite{PhysRevB.101.024303} to study the cavity-material setup, and combine it with a Hartree-Fock-like mean field treatment of the electronic interactions. 
We systematically study how Floquet engineering effects are affected by these interactions. Remarkably, rather than reducing the Floquet-induced effects, we find that cavity-mediated interactions are strongly enhanced compared to a noninteracting model with identical laser and cavity detuning to the material resonances. 
In addition, the induced interaction is strongly broadened in reciprocal space. 
We trace both the enhancement and the broadening back to the momentum space structure of the localised Frenkel excitons.  

The paper is structured as follows: in Section~\ref{sec.setup}, we introduce the model Hamiltonian and the Floquet formalism used in this manuscript. 
In Section~\ref{sec.screening}, we describe the main steps of the screening calculation. To improve the accessibility of this work, technical aspects are moved to supplementary sections. 
The effective low-energy Hamiltonian, which we obtain from the screening calculation, is then analyzed in Section~\ref{sec.screened-Hamiltonian}, where we present a detailed study of the influence of electronic interactions on the emerging Floquet physics.
We finally conclude with Section~\ref{sec.conclusions}.

\section{Setup and Model}
\label{sec.setup}

\subsection{The model}
\label{sec.model}

\textit{The static Hamiltonian--}
In this paper, we will focus on a two-dimensional electron system coupled to a single-mode cavity. The electrons are described by a two-band Hubbard model, such that the Hamiltonian of this hybrid light-electron system is given by
\begin{align}\label{eq.H0}
        \hat{H} &= \hat{h} + \hat{U}, 
\end{align}
where the single-electron Hamiltonian $\hat{h}$ of the electronic system and the cavity is given by (we set $\hbar = 1$)
\begin{align} \label{eq.h}
\hat{h}   &= \omc  \hat{a}^{\dag}\hat{a} + \sum_{{\bf{k}}, b, s} \big( \ep_{{\bf{k}}, b} -\mu \big) \hat{c}_{{\bf{k}} b, s}^{\dag} \hat{c}_{{\bf{k}} b, s} + i g_{c} \left( \hat{C} - \hat{C}^{\dag} \right)
\end{align}
with
\begin{align} \label{eq.C-def}
\hat{C} &= \frac{1 }{\sqrt{N}} \hat{a} \sum_{{\bf{k}}, s}  \hat{c}_{{\bf{k}} 2 s}^{\dag} \hat{c}_{{\bf{k}} 1 s}.
\end{align}
Here, the bosonic operator $\hat{a}$ annihilates a cavity photon with frequency $\omc$, the fermionic operator $\hat{c}_{{\bf k} b s}$ annihilates an electron with band index $b = 1,2$, spin $s$ and quasi-momentum $\bf k$, and $N$ denotes the number of lattice sites. The quasimomentum $\bf k$ is dimensionless, measured in units of the lattice constant. 

In Eq.~(\ref{eq.h}), we also added the dipole coupling defined in Eq.~(\ref{eq.C-def}) between the interband transition of the two electronic bands and the cavity (for a derivation of this coupling, see~\cite{PhysRevB.101.205140,PhysRevB.103.075131}). The cavity-electron coupling is quantified by the vacuum Rabi frequency $g_c$. 
 In the optical regime, we have $g_c \ll \omc$, which allowed us to neglect counter-rotating terms in Eq.~(\ref{eq.h}) and only include near-resonant interactions, where a photon annihilation is accompanied by an excitation of an electron into the upper band (and vice versa). 
We also introduced the chemical potential $\mu$ in Eq.~(\ref{eq.h}). Throughout this paper, we chose $\mu$ to describe a weakly doped band insulator, where the lower electron band is nearly fully filled.
In addition, electronic interactions are collected in the second term of Eq.~(\ref{eq.H0}), which reads
\begin{align} \label{eq.U-Hamiltonian}
\hat{U}   &= \hat{U}_{11} + \hat{U}_{22} + \hat{U}_{12},
\end{align}
where $\hat{U}_{bb} = U_{bb} \sum_{\bf r} \hat{n}_{{\bf r},b,\uparrow} \hat{n}_{{\bf r},b, \downarrow}$ denotes a local electron repulsion for two electrons in band $b$ both residing on lattice site ${\bf r}$. $\hat{U}_{12} = U_{12} \sum_{{\bf r}, s, s'} n_{{\bf r} 2, s} n_{{\bf r}, 1, s'}$ denotes the local repulsion between electrons residing in distinct bands. 

\textit{Laser driving--}
The static cavity-electron system is additionally driven by an external laser field with frequency $\omL$, which is propagating perpendicularly to the 2D electron system and thus couples only to vertical band transitions where the electron momentum is conserved. The laser is sufficiently detuned from the cavity resonance to directly interact with the electrons. The corresponding driving Hamiltonian is given by 
\begin{align}\label{eq.H_tot}
\hat{H}_{drive} (t) &= \hat{D} e^{-i \omL t} + \hat{D}^{\dag} e^{ i \omL t},
\end{align}
where 
\begin{equation}\label{D}
    \hat{D} = - \hat{D}^\dagger = -i \sum_{{\bf{k}}, s} ( g_{L} \hat{c}_{{\bf{k}} 2 s}^{\dag} \hat{c}_{{\bf{k}} 1 s}+\text { h.c. } )
\end{equation}
is the dipole operator with $g_L$ denoting the effective Rabi frequency of the laser drive. 
In this paper, we will consider a situation where $g_L$ is much larger than the cavity coupling of $g_c$, i.e. where the laser field amplitude is larger than the vacuum field strength of the cavity field. 
Hence, it is not apparent that counter-rotating terms can be neglected, and we retain them for now. 

\subsection{Low-energy Floquet Hamiltonians}

The full Hamiltonian of the laser-driven cavity-electron system reads $\hat{H}{(t)} = \hat{H} + \hat{H}_{drive} (t)$. 
As shown in Refs.~\cite{PhysRevA.91.033416,Giovannini_2020,PhysRevB.101.024303}, this time-dependent problem can be solved by finding the self-consistent solution of a static quasi-eigenenergy Floquet problem, 
\begin{equation}\label{0-HarmonicEigenProblem}
    \hat{H}^{{\rm Fl}}_{(\Ea)} \vert \alpha\rangle = \Ea \vert \alpha\rangle,
\end{equation}
where $\Ea$ is the Floquet quasi-eigenenergy and the effective Floquet Hamiltonian $\hat{H}^{{\rm eff}}_{(\Ea)}$ reads
\begin{equation}\label{Heff}
\begin{split}
    \hat{H}^{{\rm Fl}}_{(\Ea)} &= \hat{H} + \hat{D}^{\dag} \frac{1}{\Ea-\hat{H} +\omL-\hat{D}^{\dag} \frac{1}{\Ea-\hat{H} +2 \omL-\ldots} \hat{D}} \hat{D} \\
    & ~~~~~~~~ +
    \hat{D} \frac{1}{\Ea-\hat{H} -\omL-\hat{D} \frac{1}{\Ea-\hat{H} -2 \omL-\ldots} \hat{D}^{\dag}} \hat{D}^{\dag}.
\end{split}
\end{equation}
$\hat{H}^{{\rm Fl}}_{(\Ea)}$ is an effective Hamiltonian which acts only on the physical Hilbert space. This simplification comes at the price that $\hat{H}^{{\rm Fl}}_{(\Ea)}$ has to be determined self-consistently.

Treating the light-matter coupling perturbatively, we can expand the effective Floquet Hamiltonian (\ref{Heff}) in orders of $\hat{D}$ and $\hat{D}^\dagger$. This is appropriate, as long as the Rabi frequency $g_L$ is much smaller than the laser's detuning to material resonances. 
To the lowest order, we obtain the following Hamiltonian
\begin{equation}\label{Heff_weak_drive}
    \hat{H}^{\rm Fl}_{(\Ea)} \approx  \hat{H} + \hat{D}^{\dag}\hat{G}{(\Ea +\omL)}\hat{D} + \hat{D}\hat{G}{(\Ea -\omL)}\hat{D}^{\dag},
\end{equation}
where the Green operator (i.e. resolvent) $\hat{G}$ reads
\begin{equation}
    \hat{G}{(E)} = \frac{1}{ E - \hat{H}}.
\end{equation}
But even this simplified expression represents a correlated many-body hybrid light-matter Hamiltonian, which contains electronic interactions to arbitrary orders and which cannot be evaluated straightforwardly. 
In the literature, the high-frequency expansion \cite{Eckardt_2015,PhysRevB.93.144307} is frequently used to further simplify Eq.~ (\ref{Heff_weak_drive}). It is valid when $\vert\vert \Ea - \hat{H} \vert\vert \ll \omL$, such that the Floquet Hamiltonian is reduced to~\cite{Eckardt_2015}
\begin{equation}
\begin{split}
   \hat{H}^{\rm Fl}_{\rm high-freq } &\approx  \hat{H} + \frac{ [\hat{D}^{\dag},\hat{D}] }{\omL} +  \frac{  [\hat{D},[\hat{H},\hat{D}^{\dag}]] + [\hat{D}^{\dag},[\hat{H},\hat{D}]]  }{2 \omL^2}  + \ldots
\end{split}
\end{equation}
In this limit, the self-consistency requirement disappears, i.e., $ \hat{H}^{\rm Fl}_{\rm high-freq }$ no longer depends on $\Ea$, which greatly simplifies the eigenvalue problem (\ref{0-HarmonicEigenProblem}).
However, this high-frequency limit becomes problematic when describing in-gap driving of the two-band model considered here. 

Consequently, we solve this problem by adopting a projector technique akin to similar techniques in strongly correlated systems~\cite{essler2005one,klein1974degenerate}: As the driving will be considered off-resonant from any material resonance, we assume the self-consistent solution to Eq.~(\ref{0-HarmonicEigenProblem}) lies closely to the static (no-driving) limit. 
We thus introduce the many-body energy filter for the unperturbed Hamiltonian (see Appendix \ref{appendix-filter})
\begin{align}\label{eq.PEa}
\PEa = \delta( \Ea - \hat{H} ), 
\end{align}
and focus our attention on the low-energy limit of the projected Hamiltonian given by $\PEa \hat{H}^{\rm Fl }_{(\Ea)} \PEa$. In this limit, as we will see, the self-consistency can also be dropped and we obtain low-energy models which can be analyzed with the usual equilibrium methods. 

\section{Screening calculation and the effective Hamiltonian}
\label{sec.screening}

We now wish to evaluate the projected effective low-energy Hamiltonian 
\begin{align}
&\hat{H}_{\rm eff} (\Ea) \equiv \PEa \hat{H}^{\rm Fl}_{(\Ea)} \PEa. 
\end{align}
Inserting Eq.~(\ref{Heff_weak_drive}), we have to evaluate the action of the low-energy projector $\PEa$ on the dipole operator $\hat{D}$ followed by Green operator $\hat{G} (E)$. We will carry out this evaluation in several steps. First, as we consider a cavity-electron coupling $g_c$ which is weak compared to the bare cavity frequency $\omc$, we expand the effective Hamiltonian to its leading order in $g_c$, 
\begin{align}\label{eq.H_eff}
&\hat{H}_{\rm eff} (\Ea) \simeq \PEa \hat{H} \PEa \notag \\ 
&+ \PEa \hat{D}^\dagger \hat{G}^{(0)} (\Ea + \omL) \hat{D} \PEa + \PEa \hat{D} \hat{G}^{(0)} (\Ea - \omL) \hat{D}^\dagger \PEa \notag \\
&+ |g_c|^2 \PEa \hat{D}^\dagger \hat{G}^{(0)} (\Ea + \omL) \hat{C}  \hat{G}^{(0)} (\Ea + \omL) \hat{C}^\dagger \hat{G}^{(0)} (\Ea + \omL) \hat{D} \PEa. 
\end{align}
As we will see below, the terms in the second line will give rise to the Stokes (i.e. optical Stark) and Bloch-Siegert shifts, and the $|g_c|^2$-contribution in the third line corresponds to a cavity-mediated, laser-stimulated electron interaction. 
Any other expansion term will be suppressed relative to these leading contributions by at least $\sim g_L /\omL$ or  $\sim g_c /\omc$. 
In Eq.~(\ref{eq.H_eff}), we added the superscript ``$0$" to indicate quantities evaluated at $g_c = 0$, i.e.
\begin{align}
\hat{H}^{(0)} &= \hat{H} \big|_{g_c = 0}, \\
\hat{G}^{(0)} (E) &= \hat{G} (E) \big|_{g_c = 0}.
\end{align}
We next present a procedure to simplify the effective Hamiltonian~(\ref{eq.H_eff}) and derive a simplified Hamiltonian, where the electronic interband interactions are treated on a mean field level and give rise to excitonic screening of the Floquet Hamiltonian. 
Our strategy is as follows: 

\paragraph{Dyson equation}
First, we expand the Green operator using the Dyson equation, 
\begin{align}\label{eq.Dyson-eq}
\hat{G}^{(0)} (E) = \hat{g}^{(0)} (E) + \hat{g}^{(0)} (E) \hat{U} \hat{G}^{(0)} (E),
\end{align}
where we defined the non-interacting Green operator
\begin{align}\label{eq.g0}
\hat{g}^{(0)} (E) &\equiv \hat{G} (E) \big|_{g_c = 0, \hat{U} = 0}. 
\end{align}
We apply the Dyson equation recursively to write the interacting Green operator as
\begin{align} \label{eq.series-expansion}
\hat{G}^{(0)} (E) &= \sum\limits_{n=0}^{\infty} \left[ \hat{g}^{(0)} (E) \hat{U} \right]^n \hat{g}^{(0)} (E).
\end{align}

\paragraph{Commutators}
Second, we move the (de-)excitation operators $\hat{C}$ ($\hat{C}^\dagger$) and $\hat{D}$ ($\hat{D}^\dagger$) past the Green operators and interaction operators. Previously, this idea has been applied to study the scattering between semiconductor excitons and impurities ~\cite{combescot2004theory,piermarocchi2004coherent}. 
Following an approach pioneered by Anderson in his seminal paper on BCS theory~\cite{PhysRev.112.1900}, we only retain the contributions to the commutator, which are expected to give a finite expectation value, and neglect terms such as the intraband polarization $\langle \hat{c}^\dagger_{{\bf k}, b, s} \hat{c}_{{\bf k}', b, s} \rangle$ when ${\bf k} \neq {\bf k}'$. In particular, we use the following approximation (see Appendix~\ref{sec.commutator-relations} for the full expression)
\begin{equation}\label{U-b-commutator-reduced}
    \begin{split}
        &\hat{U} \hat{b}_{{\bf q},s}^{\dag} \approx \sum\limits_{\bf k} \hat{b}_{{\bf k},s}^{\dag} \hat{f}_{{\bf k},{\bf q}}^{s}
    \end{split}
\end{equation}
where $\hat{b}_{{\bf k},s}^{\dag} \equiv \hat{c}_{{\bf k},2,s}^{\dag} \hat{c}_{{\bf k},1,s}$ denotes the inter-band polarization operator at momentum $\bf k$ and spin $s$, which occurs in the operators $\hat{C}$ and $\hat{D}$, and
\begin{align} \label{eq.f-def}
\hat{f}_{{\bf k},{\bf q}}^{s} &= \delta_{{\bf k},{\bf q}}  
        \big( \hat{U} 
        - U_{11} \hat{\nu}_{\Bar{s}}
        + U_{12}\sum\limits_{s'} \hat{\nu}_{s'}
        \big) - \frac{U_{12}}{N} \hat{n}_{{\bf q}, 1,s}.
\end{align}
We have also defined 
the spin-resolved filling operator in the lower band
\begin{equation}\label{filling-operator}
    \hat{\nu}_{s} \equiv \frac{1}{N}\sum\limits_{{\bf k'}}\hat{n}_{{\bf k'}, 1,s}.
\end{equation}

For the non-interacting Green operator $\hat{g}$, we find exact commutation relations (see Appendix~\ref{eq.non-interacting-Green-operator}), 
\begin{equation}\label{g-commutator}
    \begin{split}
        \hat{g}^{(0)}{(E)} \hat{c}_{{\bf k} b s}^{\dag} &= \hat{c}_{{\bf k} b s}^{\dag} \hat{g}^{(0)}{(E - \ep_{{\bf k}, b} +\mu)}, \\
        \hat{g}^{(0)}{(E)} \hat{a}^{\dag} &= \hat{a}^{\dag} \hat{g}^{(0)}{(E - \omc )}.
    \end{split}
\end{equation}
Thus, for instance, in the term 
\begin{align}
\hat{D}^\dagger \hat{G}^{(0)} (E) \hat{D} = \hat{D}^\dagger \sum\limits_{n=0}^{\infty} \left[ \hat{g}^{(0)} (E) \hat{U} \right]^n \hat{g}^{(0)} (E) \hat{D}, 
\end{align}
moving the operator $\hat{D} \sim \sum_q \hat{b}_{{\bf q}, s}^{\dag}$ left will shift the argument of the noninteracting Green operators from $E$ to $E-\ep_{{\bf q},2}+\ep_{{\bf q},1}$. During this process, the commutator with each interaction vertex $\hat{U}$, according to Eq.~(\ref{U-b-commutator-reduced}), will also transform $\hat{U}$ into the operator $\hat{f}$ in Eq.~(\ref{eq.f-def}) with summations over internal momenta.

\paragraph{Low-energy restriction}
Third, we invoke a low temperature (and weak coupling) approximation, whereby the low-energy ground state does not contain cavity photons, nor upper band electrons, i.e.
\begin{equation}\label{low-energy-approx}
    \begin{split}
        \hat{a}                      \PEa &= 0   ~~~~ \text{and} ~~~~
        \hat{c}_{ {\bf k} , 2, s }   \PEa = 0    ~~~  \forall {\bf k}, s
    \end{split}. 
\end{equation}
This restricts us to an adiabatic driving regime, which does not generate cavity or electronic band excitations that may lead to heating. 
It further implies that in the undriven case cavity-induced changes of the lower electron band are neglected.

\paragraph{Resummation}
Fourth, we carry out a resummation of the series of internal interaction vertices to arrive at, e.g. (the explicit resummation is shown in Appendix~\ref{sec.resummation})
\begin{widetext}
\begin{equation}\label{eq.k-space-GRPA-Stark-middle}
    \begin{split}
        &\PEa \hat{D}^{\dag}\hat{G}^{(0)} (\Ea) \hat{D} \PEa 
        = \vert g_{L} \vert^2 \sum_s \PEa
          \sum_{{\bf k}} \hat{n}_{{\bf k}, 1, s} \hat{G}_H^s (\Ea - \Delta^0_{\bf k}) \times \left( 1 + \frac{U_{12}}{N} \sum_{\bf k'} \hat{n}_{{\bf k'}, 1, s} \hat{G}_H^s (\Ea - \Delta^0_{\bf k'}) \right)^{-1}
            \PEa, 
    \end{split}
\end{equation}
\end{widetext}
where
\begin{align}
\hat{G}_H^s (E) &= \frac{1}{ [ \hat{G}^{(0)} (E) ]^{-1} + U_{11} \hat{\nu}_{\Bar{s}} - U_{12} \sum_{s'} \hat{\nu}_{s'} }
\end{align}
and we have introduced the laser-bandgap detuning $\Delta^{0}_{\bf k}\equiv \ep_{{\bf k},2} - \ep_{{\bf k},1} - \omL$. 
The subscript "H" stands for "Hartree", since in the mean field limit below this result becomes identical to a resummation of Hartree-type Feynman diagrams. 

We then invoke the low-energy restriction again, and
use the definition of the low-energy projector in Eq.~(\ref{eq.PEa}) to write (see Appendix \ref{appendix-Corrections})
\begin{equation}\label{G1-reduced-to-number}
\begin{split}
    \hat{G}^{(0)}{( \Ea - \Delta^{0}_{\bf k})}  \PEa
    & \approx - ( \Delta^{0}_{\bf k})^{-1} \PEa,
\end{split}
\end{equation}
where $\Delta^{0}_{\bf k}$ is the laser-bandgap detuning introduced above. 
Hence, in this regime, the self-consistency condition of the Hamiltonian in Eq.~(\ref{Heff_weak_drive}) is removed, and we obtain an effective low-energy Hamiltonian which is similar - but not identical - to a high-frequency expansion. The crucial difference is that the resulting Hamiltonian contains the effective impact of electronic interactions (i.e. screening) on the low-energy degrees of freedom.

\paragraph{Mean field decoupling}
Finally, we carry out a mean field decoupling of the internal interaction vertices of the previous step, and neglect the impact of charge fluctuations on the effective Hamiltonian. This means, we replace the filling operator by its expectation value, 
\begin{equation}\label{mean-field-global}
    \hat{\nu}_s \rightarrow \nu_s = \langle \hat{\nu}_s \rangle,
\end{equation}
which is evaluated at equilibrium. Here we assume that changes to this mean field value are negligible, such that we do not need to determine it self-consistently.  
Likewise, the microscopic electronic occupation is replaced by its equilibrium expectation value,
\begin{equation}\label{mean-field-local}
    \hat{n}_{{\bf k},1, s} \rightarrow  \langle  \hat{n}_{{\bf k},1, s}  \rangle.
\end{equation}
Overall, these steps result in an effective low-temperature Hamiltonian, which we will analyze in the following section. 
In Appendix~\ref{sec.GRPA}, we have checked that our approach is equivalent to a generalized RPA calculation for near-resonant interactions.

We note that during the derivation of the semiconductor Bloch equations (SBE) \cite{lindberg1988effective}, i.e. the differential equations for the dynamics of the expectation value of $\hat{b}_{{\bf k},s}$, the approximation (\ref{U-b-commutator-reduced}) to simplify the commutator $[\hat{U},\hat{b}_{{\bf k},s}]$ is employed, too. 
However, unlike in the derivation of the SBEs where Eq.~(\ref{U-b-commutator-reduced}) is used only once before the mean-field decoupling, in our approach this commutator approximation is repeatedly used, followed by a resummation. This repeated application of Eq.~(\ref{U-b-commutator-reduced}) results in an effective Hamiltonian governing the low-energy dynamics of the lower-band electrons, which cannot be captured by SBE. We refer to Appendix \ref{Appendix:SBE-comparison} for a detailed comparison to the SBE method.
In the following, we will analyze the effective mean-field Hamiltonian, and investigate how the emergence of excitons affects the low-energy physics. 

\section{The screened effective Hamiltonian}
\label{sec.screened-Hamiltonian}

Combining the calculations discussed in Sec.~\ref{sec.screening} (for details see Appendix~\ref{sec.MF}), we derive the screened Floquet low-energy Hamiltonian for electrons in the lower electronic band as 
\begin{align}\label{eq.main-result}
\hat{H}_{\rm eff} &= \hat{h}_{\rm eff} + \hat{U}_{\rm eff},
\end{align}
with 
\begin{align}\label{eq.h_eff}
\hat{h}_{\rm eff} &= \sum_{{\bf k},  s} \big( \ep_{{\bf k}} - \frac{\vert g_{L} \vert^2}{\Delta_{{\bf k},s} } - \frac{\vert g_L \vert^2}{\Delta_{{\bf k},s}^{BS}} -\mu \big) \hat{c}_{{\bf k} s}^{\dag} \hat{c}_{{\bf k} s}
\end{align}
and
\begin{align}\label{eq.U_eff}
\hat{U}_{\rm eff} &= \hat{U}_{11}  ~ - ~ \frac{1}{N}  \sum_{\substack{{\bf k}, s \\ {\bf k'}, s'}} \frac{\vert g_L g_c \vert^2 }{ \Delta_c  \Delta_{{\bf k'},s'}   \Delta_{{\bf k},s}  } \hat{c}_{{\bf k'}, s'} ^{\dag}  \hat{c}_{{\bf k'}, s'} \hat{c}_{{\bf k}, s}^{\dag}  \hat{c}_{{\bf k}, s}.
\end{align}
To reduce clutter, we neglected the electronic band index ($b=1$) here and in the following. 
In these equations we have introduced the laser-cavity detuning $\Delta_c = \omc - \omL$, as well as the screened denominators $\Delta_{{\bf k},s} $ and $\Delta_{{\bf k},s}^{BS}$, which are defined as
\begin{widetext}
    \begin{equation}\label{renormalised-denominator}
        \begin{split}
            \Delta_{{\bf k},s} \equiv \Delta^{0}_{\bf k}- U_{11} \nu_{\Bar{s}} + U_{12} \sum\limits_{s'}\nu_{s'}
            - \frac{U_{12}}{N} \sum\limits_{ {\bf k}' }  \langle \hat{n}_{{\bf k}',s} \rangle
            \frac
            {  \Delta^{0}_{\bf k}- U_{11} \nu_{\Bar{s}}
            + U_{12} \sum\limits_{s'}\nu_{s'} }
            { \Delta^0_{\bf k'} - U_{11} \nu_{\Bar{s}}
            + U_{12} \sum\limits_{s'}\nu_{s'}},
        \end{split}
    \end{equation}
    and
    \begin{equation}\label{eq.BS-denominator}
    \begin{split}
        \Delta_{{\bf k},s}^{BS}&=
        \Delta^{0}_{\bf k}+2\omL - U_{11} \nu_{\Bar{s}}
        + U_{12} \sum\limits_{s'}\nu_{s'} - \frac{U_{12}}{N} \sum\limits_{ {\bf k}' }  \langle \hat{n}_{{\bf k}',s} \rangle
        \frac{ \Delta^{0}_{\bf k}+2\omL - U_{11} \nu_{\Bar{s}}
        + U_{12} \sum\limits_{s'}\nu_{s'} }
        { \Delta^0_{\bf k'} +2\omL - U_{11} \nu_{\Bar{s}}
        + U_{12} \sum\limits_{s'}\nu_{s'}}.
    \end{split}
\end{equation}
\end{widetext}
Here, $\Delta^{0}_{\bf k}= \ep_{{\bf k}, 2} - \ep_{{\bf k}, 1}  - \omL$ is the bare laser-bandgap detuning introduced above. 
The detunings are renormalized by the interband interaction $U_{12}$ and the intraband interaction $U_{11}$ of the lower band. Interactions in the upper band, described by $U_{22}$, are not significant in the present weak driving limit, where no real population is created in the upper band.

Let us discuss the Hamiltonian structure: The single-particle sector $\hat{h}_{\rm eff}$ contains the band dispersion $\ep_{\bf k} - \mu$ of the bare material, which is renormalised by two terms. 
We identify the first term $- |g_L|^2 / \Delta_{{\bf k},s}$ as the optical Stark shift, i.e. the laser-induced renormalization of the electronic band energy. 
In semiconductor quantum wells, this effect was first reported in Ref.~\cite{mysyrowicz1986dressed}.
The second term $- |g_L|^2 / \Delta_{{\bf k},s}^{BS}$ is the Bloch-Siegert shift in materials~\cite{PhysRev.57.522, Sie2018} which further decreases the energy of electrons in the lower band. It stems from the non-RWA terms in the laser-electron coupling. The Bloch-Siegert shift in two-level systems has been previously derived by various Floquet methods~\cite{shirley1965solution,PhysRevA.81.022117,PhysRevA.79.032301,PhysRevLett.105.257003}, our Floquet Hamiltonian method generalizes this derivation to a many-body system.

In the interaction term $\hat{U}_{\rm eff}$, the local intra-band repulsion $\hat{U}_{11}$ is not screened by projecting out the high-energy degrees of freedom - as one would expect intuitively. However, the presence of the optical cavity gives rise to a cavity-mediated laser-stimulated interaction $\sim |g_L g_c|^2 / ( \Delta_c \Delta_{\bf k} \Delta_{\bf k'})$. 
It is induced by the scattering of a laser photon into and out of the cavity via the virtual excitation of two lower band electrons into the upper band. A very similar effect was recently proposed theoretically in quantum spin systems~\cite{chiocchetta2021cavity}, and it is also closely connected to well established methods in cold atoms to generate and control long-ranged interactions~\cite{Ritsch2013, Mivehvar2021}.

We will illustrate our theory with parameters which are chosen appropriate for a tetracene-type molecular crystal~\cite{Cudazzo_2015}. This material is known to host Frenkel excitons and can be modeled by the local Hubbard-type interactions considered here\footnote{Here we ignore the local field effect which could be important, for instance, when there are several molecules in a unit cell (creating bands with exchange interactions) and the material shows strong Davydov splitting and anisotropic optical absorbance~\cite{Cudazzo_2015}.}. 
The strength of the on-site interactions are thus chosen as $U_{11}=1.6$ eV, $U_{12}=0.8$ eV.  
We further consider a simple square lattice band structure 
\begin{equation}\label{dispersion}
    \ep_{{\bf k},b} = \ep_{b} + 2 t_{b} \big(\cos(k_x) + \cos(k_y) \big),
\end{equation}
which entail the momentum-dependent bandgap 
    $\ep_{21 {\bf k}} = \ep_{21} + 2 t_{21} \big(\cos(k_x) + \cos(k_y) \big)$
with $t_{21}\equiv t_2 -t_1$ and $\ep_{21}\equiv \ep_2 - \ep_1$. 
Unless specified otherwise, the band parameters are chosen as $\ep_{21}=3.7$ eV, $t_{21}=-0.2$ eV, with $t_{1}=0.05$ eV, $t_{2}=-0.15$ eV. 
This results in a band gap which can be driven in-gap by an optical laser. We note, finally, that all of the analytical results derived below remain valid for arbitrary band-structures $\ep_{{\bf q},b}$.

In our approach, to simplify the Floquet Hamiltonian, we start from a non-interacting band-model, and carry out a resummation over RPA-type inter-band interaction vertices. This approximation is reliable (see e.g. Refs.~\cite{PhysRevB.40.3802} and \cite{haug1984electron}) only if the undriven system can be described as a Fermi liquid, which is indeed the case in our weakly hole-doped semiconductor system: the strong on-site repulsion cannot create Fermi-surface instabilities as the charge carriers (holes in the lower-band in our case) are dilute \cite{arovas2022hubbard}. More precisely, one can define the diluteness parameter $1/\ln (1/n)$, which in our case (the hole-occupancy is $n\sim5\times10^{-4}$, corresponding to a hole-density of $10^{10}\sim 10^{11} / \text{cm}^2$ in tetracene-type molecular semiconductors) evaluates to $\sim 1/8$, which is too small for Fermi surface instabilities to appear. Previous cluster perturbation simulations \cite{wang2020emergence} also agree that the Fermi liquid is a valid description for the parameters considered here up to a doping level of $n~\sim0.3$. However, we note that this diluteness is sensible to the interaction range: The diluteness parameter becomes no longer small when the interaction becomes long ranged \cite{engelbrecht1992landau}, which makes Fermi-liquid instability easier to trigger.

\subsection{The screened denominator}
\label{sec.screened-denominator}
The different terms in the screened denominator~(\ref{renormalised-denominator}) can be readily interpreted physically: 
In the absence of electron interactions, the energy required to excite a Bloch electron with quasi-momentum ${\bf k}$ is given by the energy difference $\ep_{{\bf k},2} - \ep_{{\bf k},1}$, such that the energy mismatch to the laser drive is the bare detuning $\Delta_{\bf k}^{0}$. 
This detuning is reduced by $U_{11} \nu_{\Bar{s}}$, which originates from the absence of the intra-band repulsion $\hat{U}_{11}$, once a hole is created below the Fermi surface. 
It is counteracted by $U_{12} \nu_{\Bar{s}}$ and $U_{12} \nu_{s}$. This third term in Eq.~(\ref{renormalised-denominator}) originates from the opposite (same) spin part of the inter-band repulsion $\hat{U}_{12}$ after an electron is excited to the upper band. 
The final term has the most complex structure. It is the inter-band electron-hole excitonic attraction stemming from fermionic statistics: Once a fermion with spin-momentum quantum numbers $({\bf k},s)$ is excited to the upper band, it leaves a hole at the same momentum $({\bf k},s)$. This hole counteracts the third term, but since the electrons can hop between adjacent lattice sites, the energy shift cannot completely compensate the previous effect. 

The screened detuning~(\ref{renormalised-denominator}) is also contained in the low-temperature absorbance spectrum of the bare material, where it emerges from the Kubo formula (see Appendix~\ref{sec.absorbance}). Absorbance resonances are found when the detuning vanishes, i.e. $\Delta_{{\bf k},s} = 0$. This is the case when 
\begin{equation} \label{band-resonance}
\Delta_{\bf k}^0- U_{11} \nu_{\Bar{s}} + U_{12} \sum\limits_{s'}\nu_{s'} = 0,
\end{equation}
i.e. when the laser resonantly couples to the interband transition, which is shifted by the Hartree-type mean field contribution $- U_{11} \nu_{\Bar{s}} + U_{12} \sum\limits_{s'}\nu_{s'}$,
or when 
\begin{equation}\label{exciton-resonance}
\frac{U_{12}}{N} \sum\limits_{ {\bf k}' } 
\frac{  \langle \hat{n}_{{\bf k}',s} \rangle } { \Delta_{{\bf k}'}^0 - U_{11} \nu_{\Bar{s}} + U_{12} \sum\limits_{s'}\nu_{s'}} = 1.
\end{equation}
This latter condition describes the formation of an exciton, it can also be derived from the Wannier equation (see Ref.~\cite{katsch2022excitonic,kira2006many}) for the case of on-site interactions: For a fully occupied lower-band where $\langle \hat{n}_{{\bf k},s} \rangle = \nu_{s}= \nu_{\Bar{s}}= 1$, Eq.~(\ref{exciton-resonance}) is equivalent to the semiconductor excitonic optical resonance derived in Refs.~\cite{katsch2022excitonic,kira2006many,PhysRevB.29.4401, davis1986photoemission}
\footnote{
The Wannier equation in Ref.~\cite{kira2006many}, in our on-site repulsion model with fully-occupied lower-band, becomes 
$$(\ep_{{\bf k}, 2} - \ep_{{\bf k}, 1} - U_{11}  + 2 U_{12} ) \phi({\bf k}) -\frac{U_{12}}{N}\sum\limits_{\bf k'} \phi({\bf k'}) = \omega_{ex} \phi({\bf k}) $$
which has a solution $\phi({\bf k})=(\ep_{{\bf k},2}-\ep_{{\bf k}, 1} -\omega_{ex} - U_{11}  + 2 U_{12} )^{-1}$ representing the exciton's momentum-distribution. The eigenvalue $\omega_{ex}$ satisfies
$$ \frac{U_{12}}{N}\sum\limits_{\bf k} (\ep_{{\bf k},2}-\ep_{{\bf k}, 1} -\omega_{ex} - U_{11}  + 2 U_{12} )^{-1} = 1 $$
which is identical to our result Eq.~(\ref{exciton-resonance}) when $\langle \hat{n}_{{\bf k},s} \rangle =\nu_s =1$. 
}. 
For the parameters chosen above, Eq.~(\ref{exciton-resonance}) predicts an exciton resonance at $\omega_{\text{ex}}\approx 2.71$ eV, and we will denote the laser-exciton detuning as $\Delta_{ex}\equiv \omega_{\text{ex}} - \omL$.  
Crucially, the condition Eq.~(\ref{exciton-resonance}) is independent of the momentum index $\bf k$ in $\Delta_{{\bf k},s} = 0$. This will have profound consequences for the emergent Floquet physics, as we will explore in the remainder of this manuscript.

\subsection{Bandstructure Floquet engineering}

\paragraph{Bandstructure renormalization}
\begin{figure}
\includegraphics[width=0.45\textwidth]{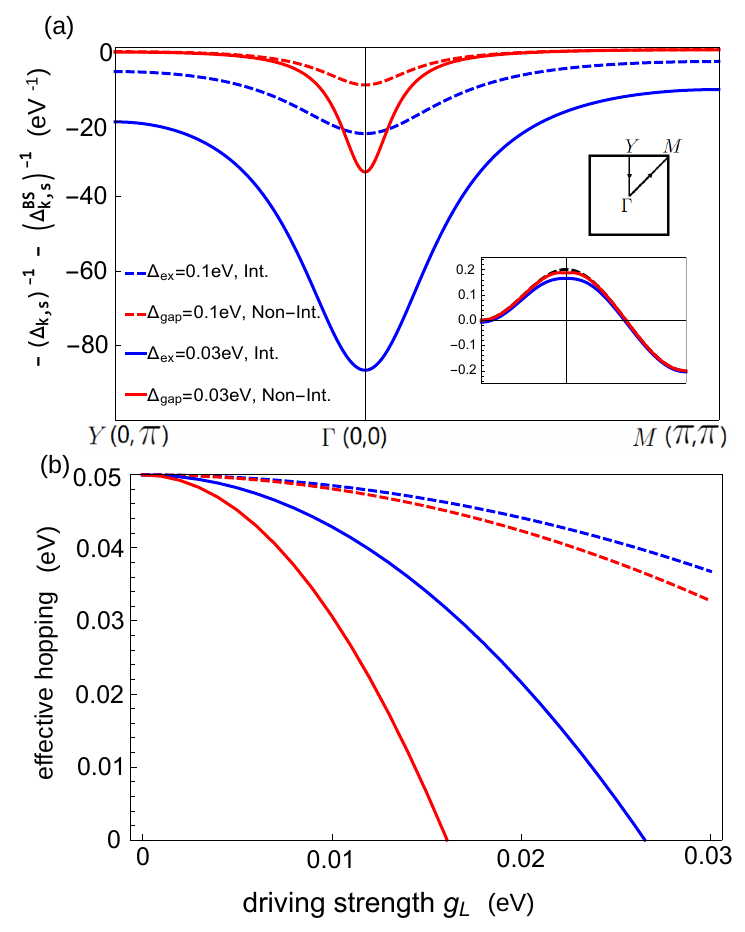}
\caption{
(a) The change of the single-electron energy of the lower-band according to Eq.~(\ref{eq.h_eff}) is shown 
along the path $Y \to \Gamma \to M$ in the first Brillouin zone. 
Since this change is proportional to $|g_L|^2$, this constant is removed. 
(b) The effective hopping rate $\Tilde{t}$, as extracted from the electronic dispersion at the $\Gamma$-point, is plotted vs. the laser driving strength $g_L$.
}
\label{fig.dispersion}
\end{figure}

We next explore how this peculiar resonance structure is reflected in the Floquet bandstructure in the single-particle sector $\hat{h}_{\rm eff}$ of Eq.~(\ref{eq.H_eff}). 
The excitonic screening behaviour in our interacting model is revealed by comparing with an unscreened case in the non-interacting model where $U_{11}=U_{12}=0$. 
To allow for a fair comparison between the screened and the unscreened cases, we fix the detuning to the respective resonances, i.e. we chose the driving frequency such that we have the same detuning to the excitonic resonance (in the screened case) and to the $\Gamma$-point of upper electron band (unscreened case). 
That is, we let $\Delta_{\text{ex}}$ in the screened case to be equal to $\Delta_{\text{gap}} \equiv \Delta_{{\bf k}=(0,0)}^0$ in the unscreened case. 
Consequently, the Stokes term $|g_L|^2 / \Delta_{{\bf k},s}$ in $\hat{h}_{\rm eff}$ has approximately the same size, and the major deviation stems from the distinct momentum dependence of the exciton resonance. 

The change of the bandstructure is shown along the high-symmetry points of the first Brillouin zone in Fig.~\ref{fig.dispersion}(a). 
Naturally, the largest renormalization effect can be achieved in the vicinity of the $\Gamma$-point for the unscreened case. The reason is obvious, as changes in different regions of reciprocal space are suppressed by a much larger detuning. 
In contrast, the screening causes changes of the bandstructure in a much larger region of reciprocal space.
This behaviour can again be explained easily: due to the lack of momentum dependence in Eq.~(\ref{exciton-resonance}), 
the detuning at momentum far from the $\Gamma$-point still vanishes as $\omL$ approaches $\omega_{\text{ex}}$. 
Hence, the screening enables the driving to renormalise the bandstructure at larger regions of reciprocal space. 
If we take $\Delta_{\text{ex}}\to0$ in Fig.~\ref{fig.dispersion}(a), the shape of the (blue) interacting curve will converge to $\phi({\bf k})$, which is the exciton's broadened momentum distribution. On the contrary, if we take $\Delta_{\text{gap}} \to 0$, the shape of the (red) non-interacting curve becomes a narrow delta peak at $\Gamma$-point. 
Naively, this would suggest an overall enhanced renormalization effect. 

In Fig.~\ref{fig.dispersion}(b), we demonstrate that this is not necessarily true: 
We plot the renormalization of the effective hopping rate $\Tilde{t}$ at the $\Gamma$-point, which we extract from the curvature of the Floquet bandstructure, versus the driving strength $g_L$.  
For a weakly hole-doped lower electron band, as we consider here, this hopping rate describes the effective kinetic energy at the Fermi surface, i.e. it is proportional to the electron mobility. 
In both screened and unscreened case, the driving leads to a reduction of the hopping rate, indicating the dynamical localisation of the mobile charge carriers due to the driving~\cite{Kuwahara2016}. 
But we find that screening reduces the renormalization of this hopping rate - even though the overall change of the bandstructure is increased. 
If we reduce the detuning to the resonance to be 0.03 eV, the unscreened calculation predicts a vanishing of the effective hopping rate at $g_L \simeq 0.015$~eV. 
This would indicate indicate a photo-induced van Hove singularity~\cite{vanHove1953}, where the density of states diverges (even though higher orders in $|g_L|$ will likely shift or destroy this singularity).
The screening counteracts the reduction of the hopping rate, such that a singularity (to leading order in $g_L$) only occurs at very large driving strengths $|g_L| \sim \Delta_{ex}$ where the weak driving approximation in Eq.~(\ref{Heff_weak_drive}) becomes very questionable and Floquet heating will become an fundamental problem.

\paragraph{Ratio between Stark and Bloch-Siegert shifts}
\begin{figure}
\includegraphics[width=0.49\textwidth]{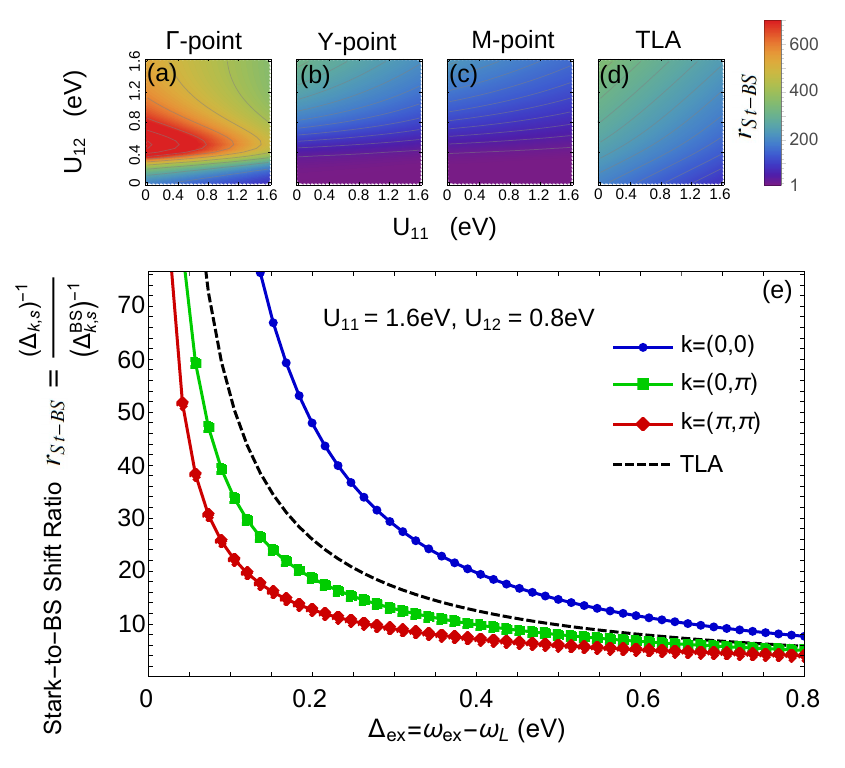}
\caption{
The ratio between the Stark shift~(\ref{renormalised-denominator}) and the BS shift~(\ref{eq.BS-denominator}) is plotted vs. the detuning between laser and exciton resonance for selected points in the Brillouin zone. For comparison, the same ratio for a two-level treatment (TLA), Eqs.~(\ref{eq.Stark-TLA}) and (\ref{eq.BS-TLA}), is shown in blue. 
The dependence of the ratio on the electronic interactions is shown at selected points in the top panels, where the laser-exciton detuning is fixed $\Delta_{\text{ex}}=0.03$eV. 
}
\label{fig.ratio}
\end{figure}
We next explore how the relative size of the Stark effect and the Bloch-Siegert shift are affected by electronic interactions. 
Recent experiments~\cite{Conway2023} on monolayer WS$_2$ revealed a enormous enhancement of the Bloch-Siegert shift compared to what one would expect from treating the exciton resonance as an effective two-level atom (TLA), where we would have
\begin{align} \label{eq.Stark-TLA}
\Delta\ep_{Stark}^{TLA} &= \frac{|g_L|^2}{\omL -\omega_{\text{ex}}},
\end{align}
and
\begin{align} \label{eq.BS-TLA}
    \Delta\ep_{BS}^{TLA} &= \frac{|g_L|^2}{\omL + \omega_{\text{ex}}},
\end{align}
where $\omega_{\text{ex}}$ is the excitonic resonance extracted from Eq.~(\ref{exciton-resonance}). 
In Fig.~\ref{fig.ratio}, we plot the ratio
\begin{align} \label{eq.r_St-BS}
r_{St-BS} &\equiv \frac{\Delta\ep_{Stark}}{\Delta\ep_{BS}}
\end{align}
vs. the detuning. We find that at the $\Gamma$-point the ratio according to Eqs.~(\ref{renormalised-denominator}) and (\ref{eq.BS-denominator}) is strongly enhanced compared to the TLA treatment in Eqs.~(\ref{eq.Stark-TLA}) and (\ref{eq.BS-TLA}). 
The excitonic enhancement benefits the Stark effect to a greater degree than the Bloch-Siegert shift, where the enhancement is suppressed by the large prefactor $2\omL$ [see Eq.~(\ref{eq.BS-denominator})]. 
However, at momenta (Y- and M-point) far from the $\Gamma$-point, our many-body approach indeed gives ratios smaller than TLA's prediction. 
If we scan the strength of the intra- and interband interactions [see the top panels of Fig.~\ref{fig.ratio}], we further find a remarkable non-monotonic dependence of this ratio at the $\Gamma$-point. 
Note that there is a weak dependence even in the TLA treatment, because we have fixed $\Delta_{\text{ex}}=0.03$eV and let $\omega_{\text{ex}}$ variate with different interaction strengths, then we cannot change $\omega_{\text{ex}}$ without changing the ratio~(\ref{eq.r_St-BS}). 
Remarkably, the ratio~(\ref{eq.r_St-BS}) displays a clear resonance-like feature, peaking at $U_{12} \simeq 0.5$~eV. 
This interesting phenomenon will be discussed in more detail in the following Section~\ref{sec.cavity-interactions}. 

A direct comparison of our results with Ref.~\cite{Conway2023} will require a simulation of the two-dimensional spectroscopy performed in this work. But as the optical signal will involve an integration over the Brillouin zone, our results already show that a many-body treatment of the excitonic resonance fundamentally can easily changes the relative size of Stark and Bloch-Siegert shifts by an order of magnitude.

\subsection{Cavity-mediated interactions}
\label{sec.cavity-interactions}
\begin{figure}
\centering
\includegraphics[width=0.45\textwidth]{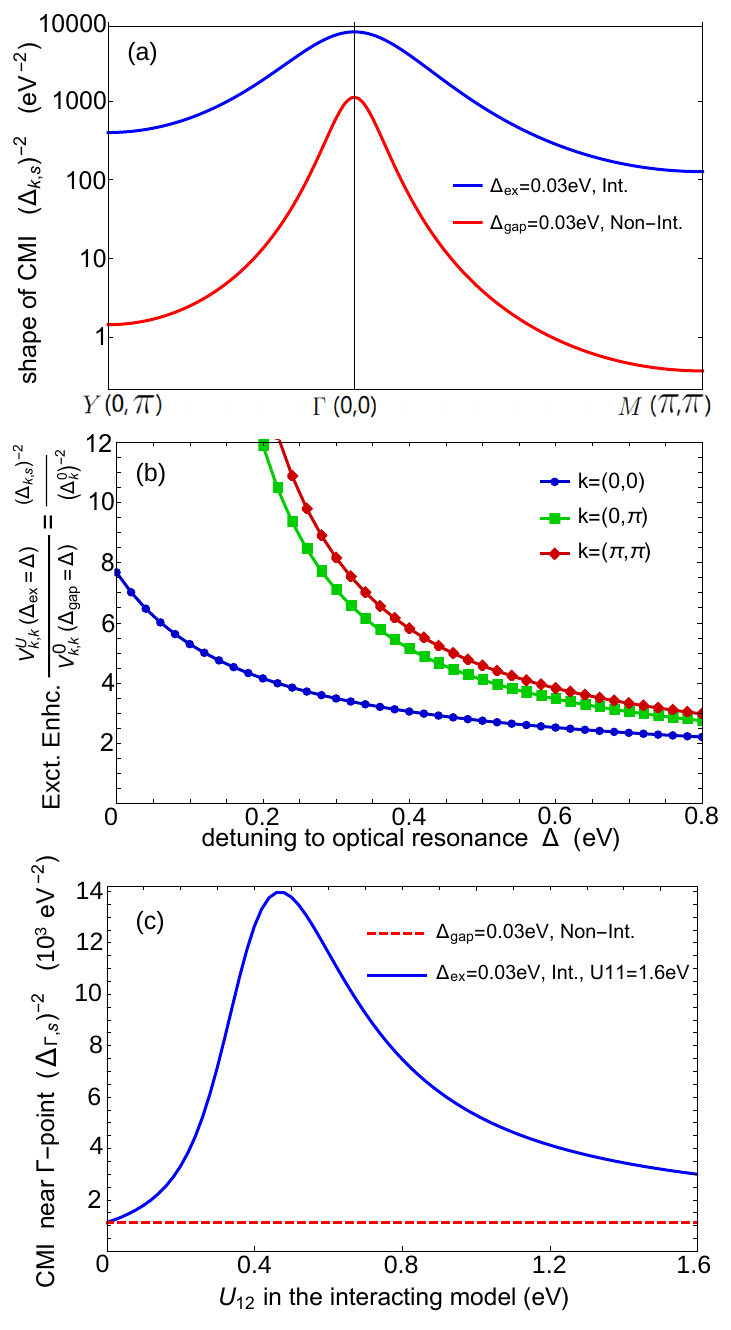}
\caption{
(a) Strength of cavity-mediated interaction in Eq.~(\ref{eq.U_eff}) in the forward scattering direction along a cut of the first Brillouin zone. The constant prefactor $|g_c g_L|^2$, which appears in Eq.~(\ref{eq.U_eff}), is removed. 
(b) The excitonic enhancement, i.e. the ratio between the cavity-mediated interaction of the interacting and the non-interacting model is plotted vs. the detuning to the respective resonance at the $\Gamma$- (${\bf k} = (0,0)$), $Y$- (${\bf k} = (0,\pi)$) and $M$-points (${\bf k} = (\pi,\pi)$).  
(c) Strength of cavity-mediated interaction in Eq.~(\ref{eq.U_eff}) at the $\Gamma$-point $(k_x, k_y) = (0, 0)$ is plotted vs. the interband interaction strength $U_{12}$. The noninteracting model is plotted as a dashed line for comparison. 
}
\label{fig.interaction}
\end{figure}

We next investigate how electronic screening affects the cavity-mediated interaction in $\hat{U}_{\rm eff}$, Eq.~(\ref{eq.U_eff}), in forward-scattering direction when ${\bf k} = {\bf k}'$, where it is proportional to $\sim 1/ \Delta_{{\bf k},s}^2$. 
In Fig.~\ref{fig.interaction}(a), we plot the inverse square of Eq.~(\ref{renormalised-denominator}) with the momentum ${\bf k}$ along the path $Y \to \Gamma \to M$. 
In addition, we provide the unscreened interaction for comparison. 
As before, in order to keep the comparison fair, we fix the detuning to the resonance (i.e. either to the exciton or to the upper electronic band). 
Remarkably, we find that the screening drastically \textit{enhances} the cavity-mediated interaction. 
This is clearly seen in Fig.~\ref{fig.interaction}(b), where we depict the ratio between the two cases as a function of the detuning to the respective resonances. 
The enhancement decreases with increasing detuning, but remains finite even for large detunings of half an electronvolt. 

This naturally leads us to investigate optimal conditions to maximize the excitonic enhancement, which we illustrate in Fig.~\ref{fig.interaction}(c). 
At a fixed detuning $\Delta = 0.05$~eV and fixed intraband repulsion $U_{11} = 1.6$~eV, we find a pronounced peak of the excitonic enhancement at $U_{12} \simeq 0.5$~eV. 
In this optimal regime, a massive enhancement of a factor $\gtrsim 10$ relative to the noninteracting model with the same detuning and driving strength is observed.
This enhancement is tied to the emergence of the exciton resonance in Eq.~(\ref{exciton-resonance}), and therefore the mixing of electronic momenta via the scattering of the virtual exciton. It has a non-trivial dependence on electron dispersion: 
If we neglect the electronic band dispersion and replace $\ep_{{\bf k} 21}$ in Eq.~(\ref{exciton-resonance}) by the constant detuning $\ep_{21}$, the exciton binding energy is simply given by $U_{12}$. 
However, as this exciton does not have a momentum structure, which is distinct from the electrons, the enhancement vanishes in this case and we recover the same interaction strength as the non-interacting model. 
This dependence on the electron dispersion makes it difficult to find a simple analytical expression for the optimal detuning. Nevertheless, with the help of the effective Hamiltonian~(\ref{eq.H_eff}) one can straightforwardly optimise the laser detuning to maximize the possible enhancement for a given material.

\subsection{Discussion}

\begin{figure}[t]
\centering
\includegraphics[width=0.49\textwidth]{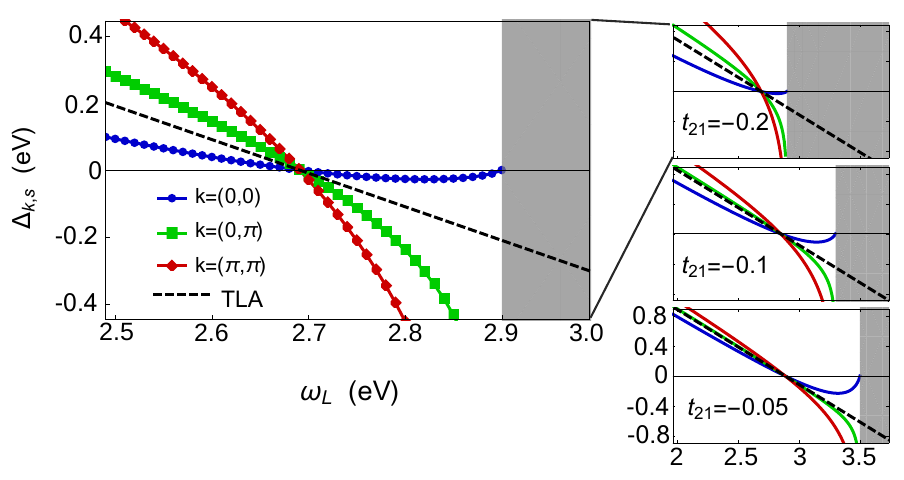}
\caption{The screened denominator~(\ref{renormalised-denominator}) is plotted vs. the driving frequency $\omL$. The model parameters are given in Section~\ref{sec.model}, with $t_1$ changed to obtain $t_{21}$ as indicated in the panels. 
The black dashed line indicates the detuning in the non-interacting model at the $\Gamma$-point with $\Delta_{\text{gap}}=\Delta_{\text{ex}}$, which is equivalent to the two-level treatment of the Stark shift in Eq.~(\ref{eq.Stark-TLA}). 
    The shaded region represents the upper electronic band, where the effective Hamiltonian~(\ref{eq.H_eff}) diverges due to resonant coupling. 
    }
\label{fig.denominator}
\end{figure}

We finish with a discussion of the excitonic resonance structure, which we already touched upon in Section~\ref{sec.screened-denominator}. 
Evidently, all our results can be traced back to the momentum structure of the detuning~(\ref{renormalised-denominator}). 
Thus, we plot the denominator~(\ref{renormalised-denominator}) in Fig.~\ref{fig.denominator} as a function of the driving frequency $\omL$ at several points in the Brillouin zone. 

In the vicinity of the $\Gamma$-point (blue line), the detuning is strongly reduced relative to the TLA treatment, which gives rise the observed enhancement of the cavity-mediated interaction. 
When the bandwidth of the lower band is reduced in the smaller panels on the right, this reduction is lost, and the detuning (at every point in the Brillouin zone) approaches the TLA.  

Moreover, Fig.~\ref{fig.denominator} also illustrates the origin of the broadening of the interaction: the detuning vanishes at the excitonic resonances \textit{for all momenta}. 
In contrast, at the band edge, the detuning only ever vanishes at a single resonant momentum, while the remaining momenta are off-resonant. 
This is what creates the broadening of the Floquet renormalization of the band structure, which we observed in Fig.~\ref{fig.dispersion}, as well as of the broadening of the cavity-mediated interaction in Fig.~\ref{fig.interaction}.

\section{Conclusions}
\label{sec.conclusions}
In this work, we investigated the electronic screening Floquet engineering effects and cavity-mediated interactions. 
We showed how exciton formation due to electronic interactions enhances these driving-induced effects by carrying out an inter-band screening calculation in combination with a mean-field decoupling of the interactions.  
Altogether, we find that the screened Floquet Hamiltonian looks superficially similar to one of a noninteracting system (which we obtain readily from Eq.~(\ref{eq.main-result}) by simply setting $U_{11} = U_{12} = 0$). 
However, the Floquet-induced change of the bandstructure is enhanced by the interaction across much of reciprocal space.
Both the electronic dispersion and the cavity-mediated interactions are strongly changed in both amplitude and range: 
In the direct vicinity of the $\Gamma$-point, the dynamical localisation due to Stark and Bloch-Siegert shifts is reduced, and their relative strength is shown to depend on the electronic interaction strengths. 
Additionally, cavity-mediated electronic interactions in the conduction band are enhanced by up to one order of magnitude and broadened in reciprocal space, such that the enhancement is even greater in other parts of the Brillouin zone. 
Our observations are readily explained by the mixing of momenta associated with the exciton, which introduces a dispersionless resonance in the Hamiltonian, and which can be targeted by an optimized choice of driving parameters from any point in reciprocal space, and thus gives rise to a strong enhancement of the interactions. 
They are reminiscent of the well established Coulomb enhancement of light-matter coupling in 2D semiconductors~\cite{Wang2018,RevModPhys.82.1489}. 

Our results show how it is possible to incorporate static screening effects into effective low-energy Floquet Hamiltonians without the need to use the rotating wave approximation. The present Hartree-type screening calculation could be extended to account for other many-body orders in the bare ground state of a correlated material. Going beyond the low-energy projection we used in our present work, it will be interesting to analyze these screening effects in stronger-driving or ultrastrong cavity-coupling regimes, where the present approximations seize to be valid and a self-consistent evaluation of the effective Hamiltonian will be essential. 
Furthermore, we have focused our attention on Frenkel excitons, which emerge from local interactions. It will be interesting to extend this method to Wannier excitons, and investigate how the excitonic enhancement is related to the screening of the Coulomb interaction in materials. This will also be necessary to allow for quantitative comparisons with recent experiments in dichalcogenides~\cite{Conway2023} as well as for future optical control applications which exploit the coupling to cavities, such as the proposed cavity quantum spin liquids~ \cite{chiocchetta2021cavity}. 

As we are considering a near fully occupied lower band, the doublon density therein is almost 1 per-site and the holon density can be ignored, and thus we mainly have the inter-band excitonic excitation. When the hole doping in the lower band becomes higher (e.g. $\sim10^{12}/\text{cm}^2$ in tetracene-type, or $\sim10^{13}/\text{cm}^2$ in TMDC materials, see Refs below), the charge-carriers (holes) in the lower band will begin interacting with the excitonic excitation, and intraband excitations will become relevant. This will result in many highly nontrivial effects: creating trion excitation resonances and exciton-trion polaritons \cite{zhumagulov2022microscopic}, smoothing the excitonic resonant peak in the absorption spectrum \cite{gao2016dynamical}, creating exciton-hole Auger recombination \cite{chen2021approaching}, ect. In the extreme doping limit where the lower band becomes half-filled, it will be very interesting to study how the doublon-holon pairs will interact with the inter-band excitations.

\textit{Acknowledgment.} 
X. W. thanks Jiajun Li for inspiring discussions. The authors acknowledge support from the Cluster of Excellence 'Advanced Imaging of Matter' of the Deutsche Forschungsgemeinschaft (DFG) - EXC 2056 - project ID 390715994.

\bibliography{cavityBCS}

\appendix

\section{Many-body projector}
\label{appendix-filter}
According to Eq.~(\ref{Heff_weak_drive}), the Floquet Hamiltonian has the form $ \hat{H}^{{\rm Fl}}_{(\Ea)} = \hat{H} + \vert g_L \vert^2 \hat{V}_{(\Ea)} $, where the static Hamiltonian $\hat{H}$ represents the unperturbed system, and the perturbation operator $\hat{V}_{(\Ea)}$ becomes independent of driving strength. The effective Hamiltonian of $\hat{H}^{{\rm Fl}}_{(\Ea)}$ near eigenenergy $\Ea$, according to the second order Kato's expansion in Refs.~\cite{essler2005one,klein1974degenerate}, is given by
\begin{equation}
\hat{H}_{\rm eff} (\Ea) = \PEa \big(   \hat{H}  + \vert g_L \vert^2 \hat{V}_{(\Ea)}  +   \vert g_L \vert^4  \hat{V}_{(\Ea)}  \sum\limits_{E_{\beta} \neq \Ea} \frac{\hat{\mathcal{P}}_{E_{\beta}}}{ \Ea - E_{\beta} }  \hat{V}_{(\Ea)}     \big) \PEa
\end{equation}
where $\PEa = \delta(\Ea-\hat{H})$ is the energy projector of the unperturbed system. 
Consistent with the weak-driving approximation made in Eq.~(\ref{Heff_weak_drive}), we only retain to the $\vert g_L \vert^2$ order, thus we get $\hat{H}_{\rm eff} (\Ea) = \PEa \hat{H}^{{\rm Fl}}_{(\Ea)} \PEa$, which is the starting point of Section \ref{sec.screening}.

\section{Commutator relations}
\label{sec.commutator-relations}

\subsection{Exact interacting commutator}

The commutation relation in Eq.~(\ref{U-b-commutator-reduced}) of the main text is our central approximation. It discards all non-commuting terms other than the Hartree-/Fock-type terms. 
We justify it in detail here: Using the definition~(\ref{eq.U-Hamiltonian}) of the interaction term $\hat{U}=\hat{U}_{11}+\hat{U}_{12}+\hat{U}_{22}$ in momentum space, and neglecting the Umklapp scattering terms which is consistent with ignoring the local field effect (i.e. the dipole-dipole interaction)~\cite{PhysRevB.29.4401,PhysRev.129.62}, we have
\begin{widetext}
\begin{equation}\label{U-b-commutator-appendix}
    \begin{split}
        &\hat{U} \hat{c}_{\mathbf{q} 2 s}^{\dag} \hat{c}_{\mathbf{q} 1 s}
        = \hat{c}_{\mathbf{q} 2 s}^{\dag} \hat{c}_{\mathbf{q} 1 s} \hat{U} + \big[ \hat{U}_{11}, \hat{c}_{\mathbf{q} 2 s}^{\dag} \hat{c}_{\mathbf{q} 1 s} \big] + \big[ \hat{U}_{12}, \hat{c}_{\mathbf{q} 2 s}^{\dag} \hat{c}_{\mathbf{q} 1 s} \big] + \big[ \hat{U}_{22}, \hat{c}_{\mathbf{q} 2 s}^{\dag} \hat{c}_{\mathbf{q} 1 s} \big] \\
        &= \hat{c}_{\mathbf{q} 2 s}^{\dag} \hat{c}_{\mathbf{q} 1 s} \hat{U} - 
           2\frac{U_{11}}{2N} \hat{c}_{\mathbf{q} 2 s}^{\dag} \sum\limits_{{\bf k'},{\bf q'}} \hat{c}_{{\bf k'}-{\bf q'}, 1, \Bar{s}}^{\dag} \hat{c}_{{\bf k'}, 1, \Bar{s}} \hat{c}_{{\bf q}-{\bf q'}, 1, s}  \\
        &~~ -\frac{U_{12}}{N} \sum\limits_{\bf q'} \hat{c}_{{\bf q}+{\bf q'}, 2, s}^{\dag} \hat{c}_{{\bf q}+{\bf q'}, 1, s} \\
        &~~ + \frac{U_{12}}{N} \sum\limits_{{\bf k'},{\bf q'},s'} \hat{c}_{{\bf q}+{\bf q'}, 2, s }^{\dag} \hat{c}_{{\bf q}, 1, s} \hat{c}_{{\bf k'}-{\bf q'}, 1, s'}^{\dag} \hat{c}_{{\bf k'}, 1, s'} 
        - \hat{c}_{{\bf q}, 2, s}^{\dag} \hat{c}_{{\bf k'}+{\bf q'}, 2, s'}^{\dag} \hat{c}_{{\bf k'}, 2, s'} \hat{c}_{{\bf q}+{\bf q'}, 1, s} \\
        &~~ - 2\frac{U_{22}}{2N}  \sum\limits_{{\bf k'},{\bf q'} } \hat{c}_{{\bf k'}, 2, \Bar{s}}^{\dag} \hat{c}_{{\bf q}-{\bf q'}, 2, s}^{\dag}  \hat{c}_{{\bf k'}-{\bf q'}, 2, \Bar{s}} \hat{c}_{\mathbf{q} 1 s} \\
        &= \hat{c}_{\mathbf{q} 2 s}^{\dag} \hat{c}_{\mathbf{q} 1 s} 
        \big( \hat{U} 
        - \frac{U_{11}}{N}\sum\limits_{\bf k}\hat{n}_{{\bf k},1,\Bar{s}} 
        + \frac{U_{12}}{N}\sum\limits_{{\bf k},s'}\hat{n}_{{\bf k},1,s'}
        \big) 
        -  \frac{ U_{12} }{N} 
        \big(
        \sum\limits_{\bf k} \hat{c}_{{\bf k} 2 s}^{\dag} \hat{c}_{{\bf k} 1 s}
        \big)
        \hat{n}_{{\bf q},1,s} \\
        & ~~ + ~~ ...
    \end{split}
\end{equation}
where $\Bar{s}=-s$. In the last line we keep only three terms: 
\end{widetext}
1) In $\big[ \hat{U}_{11}, \hat{c}_{\mathbf{q} 2 s}^{\dag} \hat{c}_{\mathbf{q} 1 s} \big]$, we only keep ${\bf q'}=0$ term
\begin{align}
    -&\frac{U_{11}}{N} \hat{c}_{\mathbf{q} 2 s}^{\dag} \sum\limits_{{\bf k'},{\bf q'}=0} \hat{c}_{{\bf k'}-{\bf q'}, 1, \Bar{s}}^{\dag} \hat{c}_{{\bf k'}, 1, \Bar{s}} \hat{c}_{{\bf q}-{\bf q'}, 1, s} \notag \\
    &= ~
    -\frac{U_{11}}{N} \hat{c}_{\mathbf{q} 2 s}^{\dag} \hat{c}_{{\bf q}, 1, s} \sum\limits_{{\bf k'}} \hat{n}_{{\bf k'}, 1, \Bar{s}} 
\end{align}
corresponding to the intra-band Hartree term. The semi-classical justification for this treatment is as follows: When ${\bf q'}\neq0$, the expectation value $\sum_{\bf k'} \langle \hat{c}_{{\bf k'}-{\bf q'}, 1, \Bar{s}}^{\dag} \hat{c}_{{\bf k'}, 1, \Bar{s}} \rangle$ represents the charge density wave of the lower-band electron with spin $\Bar{s}$ at wave-vector $\bf q'$. However, this expectation value becomes much stronger at ${\bf q'}=0$, because in this case the expectation value $\sum_{\bf k'} \langle \hat{c}_{{\bf k'}, 1, \Bar{s}}^{\dag} \hat{c}_{{\bf k'}, 1, \Bar{s}} \rangle$ represents the total electron number in the lower-band with spin $\Bar{s}$. Thus when the electron occupancy approaches 1, we can ignore all terms in $\big[ \hat{U}_{11}, \hat{c}_{\mathbf{q} 2 s}^{\dag} \hat{c}_{\mathbf{q} 1 s} \big]$ except for the ${\bf q'}=0$ contribution.

2) In $\big[ \hat{U}_{12}, \hat{c}_{\mathbf{q} 2 s}^{\dag} \hat{c}_{\mathbf{q} 1 s} \big]$, we only keep two terms: the first term is the inter-band Hartree term where (as justified above) we take ${\bf q'}=0 $ in the following summation,
\begin{align}
    &\frac{U_{12}}{N} \sum\limits_{{\bf k'},{\bf q'}=0,s'} \hat{c}_{{\bf q}+{\bf q'}, 2, s }^{\dag} \hat{c}_{{\bf q}, 1, s} \hat{c}_{{\bf k'}-{\bf q'}, 1, s'}^{\dag} \hat{c}_{{\bf k'}, 1, s'} ~ \notag \\
    &= ~
     \frac{U_{12}}{N} \hat{c}_{\mathbf{q} 2 s}^{\dag} \hat{c}_{{\bf q}, 1, s} \sum\limits_{{\bf k'},s'} \hat{n}_{{\bf k'}, 1, s'}.
\end{align}
The second term is the inter-band Fock term, where we take ${\bf q'}={\bf k'}-{\bf q} $ in the following part of $\big[ \hat{U}_{12}, \hat{c}_{\mathbf{q} 2 s}^{\dag} \hat{c}_{\mathbf{q} 1 s} \big]$
\begin{equation*}
\begin{split}
    & -\frac{U_{12}}{N} \sum\limits_{\bf q'} \hat{c}_{{\bf q}+{\bf q'}, 2, s}^{\dag} \hat{c}_{{\bf q}+{\bf q'}, 1, s} + \frac{U_{12}}{N} \sum\limits_{{\bf q'}={\bf k'}-{\bf q} \atop {\bf k'},s'=s} \hat{c}_{{\bf q}+{\bf q'}, 2, s }^{\dag} \hat{c}_{{\bf q}, 1, s} \hat{c}_{{\bf k'}-{\bf q'}, 1, s'}^{\dag} \hat{c}_{{\bf k'}, 1, s'} \\
    &= -\frac{U_{12}}{N} \sum\limits_{\bf q'} \hat{c}_{{\bf q}+{\bf q'}, 2, s}^{\dag} \hat{c}_{{\bf q}+{\bf q'}, 1, s} + \frac{U_{12}}{N} \sum\limits_{{\bf k'}} \hat{c}_{{\bf k'}, 2, s }^{\dag} (1-\hat{n}_{{\bf q}, 1, s}) \hat{c}_{{\bf k'}, 1, s} \\
    &= -\frac{U_{12}}{N}  \sum\limits_{{\bf k'}} \hat{c}_{{\bf k'}, 2, s }^{\dag} \hat{n}_{{\bf q}, 1, s}  \hat{c}_{{\bf k'}, 1, s}  \\
    &=  -\frac{U_{12}}{N} \big( \sum\limits_{{\bf k}} \hat{c}_{{\bf k}, 2, s }^{\dag} \hat{c}_{{\bf k}, 1, s} \big) \hat{n}_{{\bf q}, 1, s} ~ + ~ \frac{U_{12}}{N} \hat{c}_{{\bf q}, 2, s }^{\dag} \hat{c}_{{\bf q}, 1, s} \\
    &\approx  -\frac{U_{12}}{N} \big( \sum\limits_{{\bf k}} \hat{c}_{{\bf k}, 2, s }^{\dag} \hat{c}_{{\bf k}, 1, s} \big) \hat{n}_{{\bf q}, 1, s}  ~~ \text{when the electron number} \gg 1
\end{split}
\end{equation*}
In the last line we discard the term $\frac{U_{12}}{N} \hat{c}_{{\bf q}, 2, s }^{\dag} \hat{c}_{{\bf q}, 1, s}$, as it will be much smaller than the inter-band Hartree term $-\frac{U_{11}}{N} \hat{c}_{\mathbf{q} 2 s}^{\dag} \hat{c}_{{\bf q}, 1, s} \sum_{{\bf k'}} \hat{n}_{{\bf k'}, 1, \Bar{s}}  $ (which has the same form), when the total electron number in the lower-band with spin $\bar{s}$ is large. The semi-classical justification for only keeping the ${\bf q'}={\bf k'}-{\bf q} \And s'=s$ part in the above summation is that, when ignoring the electron correlation in the lower-band, the expectation value will vanish, i.e. $\langle \hat{c}_{{\bf q}, 1, s} \hat{c}_{{\bf k'}-{\bf q'}, 1, s'}^{\dag} \rangle = 0$, unless ${\bf q'}={\bf k'}-{\bf q} \And s'=s$. 

In the commutator $\big[ \hat{U}_{12}, \hat{c}_{\mathbf{q} 2 s}^{\dag} \hat{c}_{\mathbf{q} 1 s} \big]$, we discard the last term 
\begin{equation*}
    \hat{c}_{{\bf q}, 2, s}^{\dag} \hat{c}_{{\bf k'}+{\bf q'}, 2, s'}^{\dag} \hat{c}_{{\bf k'}, 2, s'} \hat{c}_{{\bf q}+{\bf q'}, 1, s}
\end{equation*}
because this term vanishes in the low-energy limit where $\hat{c}_{{\bf k'}, 2, s'} \PEa =0$.

3) For the same reason, the commutator $\big[ \hat{U}_{22}, \hat{c}_{\mathbf{q} 2 s}^{\dag} \hat{c}_{\mathbf{q} 1 s} \big]$ can be ignored in the low-energy limit. 

Altogether, in the last line of Eq.~(\ref{U-b-commutator-appendix}), we keep the intra-band Hartree term, inter-band Hartree term and the inter-band Fock term of the commutator $\big[ \hat{U}, \hat{c}_{\mathbf{q} 2 s}^{\dag} \hat{c}_{\mathbf{q} 1 s} \big]$. The approximation made in Eq.~(\ref{U-b-commutator-appendix}) is analogous to the random phase approximation in the equation-of-motion method, as elucidated in Ref.~\cite{rowe1968equations}, where the commutator terms are discarded if their expectation value in the non-interacting ground state vanishes.

\subsection{Non-interacting Green operator}
\label{eq.non-interacting-Green-operator}

\subsubsection{Proving $\hat{g}^{(0)}{(E)} \hat{c}_{{\bf k} b s}^{\dag} = \hat{c}_{{\bf k} b s}^{\dag} \hat{g}^{(0)}{(E - \ep_{{\bf k}, b} +\mu)} $}
We first evaluate $\hat{g}^{(0)}{(\omega)} \hat{c}_{\mathbf{q} b s}^{\dag}$ for an arbitrary electron quasi-momentum $\bf q$, band-index $b$ and spin $s$. The non-interacting Green operator $\hat{g}^{(0)}{(\omega)}$ is defined in Eq.~(\ref{eq.g0}). It is the resolvent of the following free Hamiltonian $\hat{h}^{(0)}$, defined as 
\begin{equation}
    \hat{h}^{(0)} = \hat{h} \big|_{g_c = 0} 
    = \omc  \hat{a}^{\dag}\hat{a} + \sum_{{\bf{k}}, b, s} \big( \ep_{{\bf{k}}, b} -\mu \big) \hat{c}_{{\bf{k}} b, s}^{\dag} \hat{c}_{{\bf{k}} b, s}.
\end{equation}
$\hat{h}^{(0)}$ commutes with the kinetic Hamiltonian $\hat{H}_{{\bf q},b ,s}$, $ [ \hat{h}^{(0)} , \hat{H}_{{\bf q},b ,s} ] =0  $, where we define 
\begin{equation}\label{def-kinetic-Hamiltonian}
    \hat{H}_{{\bf q}, b,s} \equiv  (\ep_{\mathbf{q}, b} -\mu ) \hat{c}_{\mathbf{q} b s}^{\dag} \hat{c}_{\mathbf{q} b s}.
\end{equation}
This allows us to write down the following simplified Dyson expansion
\begin{equation}
    \begin{split}
        \hat{g}^{(0)}{(\omega)} &= \frac{1}{\omega - (\hat{h}^{(0)} - \hat{H}_{{\bf q}, b,s} ) - \hat{H}_{{\bf q}, b,s} } \\
        &= \sum\limits_{n=0}^{\infty} (\frac{1}{\omega - (\hat{h}^{(0)} - \hat{H}_{{\bf q}, b,s} )})^{n+1} (\hat{H}_{{\bf q}, b,s})^{n}
    \end{split}
\end{equation}
and therefore 
\begin{equation}
    \begin{split}
        \hat{g}^{(0)}{(x)} \hat{c}_{\mathbf{q} b s}^{\dag}  &= \sum\limits_{n=0}^{\infty} (\frac{1}{\omega - (\hat{h}^{(0)} - \hat{H}_{{\bf q}, b,s} )})^{n+1} (\epsilon_{\mathbf{q}, b} -\mu )^{n}  (\hat{c}_{\mathbf{q} b s}^{\dag} \hat{c}_{\mathbf{q} b s})^{n} \hat{c}_{\mathbf{q} b s}^{\dag} \\
        &= \sum\limits_{n=0}^{\infty} (\frac{1}{\omega - (\hat{h}^{(0)} - \hat{H}_{{\bf q}, b,s} )})^{n+1} (\epsilon_{\mathbf{q}, b} -\mu )^{n}  \hat{c}_{\mathbf{q} b s}^{\dag} (\hat{c}_{\mathbf{q} b s} \hat{c}_{\mathbf{q} b s}^{\dag} )^{n} \\
        &= \sum\limits_{n=0}^{\infty} (\frac{1}{\omega - (\hat{h}^{(0)} - \hat{H}_{{\bf q}, b,s} )})^{n+1} (\epsilon_{\mathbf{q}, b} -\mu )^{n}  \hat{c}_{\mathbf{q} b s}^{\dag} (1 - \hat{c}_{\mathbf{q} b s}^{\dag} \hat{c}_{\mathbf{q} b s} )^{n} \\
        &= \hat{c}_{\mathbf{q} b s}^{\dag} \sum\limits_{n=0}^{\infty} (\frac{1}{\omega - (\hat{h}^{(0)} - \hat{H}_{{\bf q}, b,s} )})^{n+1} (\epsilon_{\mathbf{q}, b} -\mu )^{n}   (1 - \hat{c}_{\mathbf{q} b s}^{\dag} \hat{c}_{\mathbf{q} b s} )^{n} \\
        &= \hat{c}_{\mathbf{q} b s}^{\dag} \sum\limits_{n=0}^{\infty} (\frac{1}{\omega - (\hat{h}^{(0)} - \hat{H}_{{\bf q}, b,s} )})^{n+1} (\epsilon_{\mathbf{q}, b} -\mu  - \hat{H}_{{\bf q}, b,s} )^{n}.
    \end{split}
\end{equation}
In the second line we just shifted the position of $n$ parentheses without moving any operator, in the third line we use the fermionic commutation relation $ \{\hat{c}_{\mathbf{q} b s}, \hat{c}_{\mathbf{q} b s}^{\dag} \}=1$. In the fourth line, we use the fact that $\hat{H}_{{\bf q},b ,s}$ is the only part in $\hat{h}^{(0)}$ that acts non-trivially on $\hat{c}_{\mathbf{q} b s}^{\dag}$, i.e. we use $ [ \hat{h}^{(0)} - \hat{H}_{{\bf q},b ,s} , ~ \hat{c}_{\mathbf{q} b s}^{\dag} ] =0  $ to move $\hat{c}_{\mathbf{q} b s}^{\dag}$ to the very left. In the fifth (last) line we use the definition of $\hat{H}_{{\bf q}, b,s}$ in Eq.~(\ref{def-kinetic-Hamiltonian}).

However, since $ [ \hat{h}^{(0)} , \hat{H}_{{\bf q},b ,s} ] =0  $, we can reverse the following Dyson expansion
\begin{equation}
    \begin{split}
        &\sum\limits_{n=0}^{\infty} (\frac{1}{\omega - (\hat{h}^{(0)} - \hat{H}_{{\bf q}, b,s} )})^{n+1} (\epsilon_{\mathbf{q}, b} -\mu  - \hat{H}_{{\bf q}, b,s} )^{n} \\
        &=  \frac{1}{\omega - (\hat{h}^{(0)} - \hat{H}_{{\bf q}, b,s} ) - (\epsilon_{\mathbf{q}, b} -\mu  - \hat{H}_{{\bf q}, b,s} ) }  \\
        &= \frac{1}{(\omega -\epsilon_{\mathbf{q}, b} +\mu  ) - \hat{h}^{(0)}   } \\
        &= \hat{g}^{(0)}{(\omega -\epsilon_{\mathbf{q}, b} +\mu )}
    \end{split}
\end{equation}
Thus we derive the exact equation in Eq.~(\ref{g-commutator}) which reads
\begin{equation}
    \hat{g}^{(0)}{(\omega)} \hat{c}_{\mathbf{q} b s}^{\dag} = \hat{c}_{\mathbf{q} b s}^{\dag} \hat{g}^{(0)}{(\omega -\epsilon_{\mathbf{q}, b} +\mu)}. \label{eq.B6}
\end{equation}

Taking the Hermitian conjugate of both sides, and then shifting the argument $\omega \to \omega + \epsilon_{\mathbf{q}, b} -\mu $, we get
\begin{equation}
    \hat{g}^{(0)}{(\omega)} \hat{c}_{\mathbf{q} b s} = \hat{c}_{\mathbf{q} b s} \hat{g}^{(0)}{(\omega + \epsilon_{\mathbf{q}, b} -\mu)}. \label{eq.B7}
\end{equation}

\subsubsection{Proving $\hat{g}^{(0)}{(E)} \hat{a}^{\dag} = \hat{a}^{\dag} \hat{g}^{(0)}{(E - \omc )}$}
Similarly, we next evaluate $\hat{g}^{(0)}{(\omega)} \hat{a}^{\dag} $, where this time $\hat{a}^{\dag}$ is a bosonic operator creating a cavity photon. Since $ [ \hat{h}^{(0)} , \omega_{c} \hat{a}^{\dag}\hat{a}  ] =0  $, we have the following simplified Dyson expansion
\begin{equation}
    \begin{split}
        \hat{g}^{(0)}{(\omega)} &= \frac{1}{\omega - (\hat{h}^{(0)} - \omega_{c} \hat{a}^{\dag}\hat{a} ) - \omega_{c} \hat{a}^{\dag}\hat{a} } \\
        &= \sum\limits_{n=0}^{\infty} (\frac{1}{\omega - (\hat{h}^{(0)} - \omega_{c} \hat{a}^{\dag}\hat{a} )})^{n+1} (\omega_{c} \hat{a}^{\dag}\hat{a})^{n},
    \end{split}
\end{equation}
and thus
\begin{equation}
    \begin{split}
        \hat{g}^{(0)}{(\omega)} \hat{a}^{\dag}
        &= \sum\limits_{n=0}^{\infty} (\frac{1}{\omega - (\hat{h}^{(0)} - \omega_{c} \hat{a}^{\dag}\hat{a} )})^{n+1} (\omega_{c})^{n} (\hat{a}^{\dag}\hat{a})^{n} \hat{a}^{\dag} \\
        &= \sum\limits_{n=0}^{\infty} (\frac{1}{\omega - (\hat{h}^{(0)} - \omega_{c} \hat{a}^{\dag}\hat{a} )})^{n+1} (\omega_{c})^{n} \hat{a}^{\dag}(\hat{a} \hat{a}^{\dag})^{n}\\
        &= \sum\limits_{n=0}^{\infty} (\frac{1}{\omega - (\hat{h}^{(0)} - \omega_{c} \hat{a}^{\dag}\hat{a} )})^{n+1} (\omega_{c})^{n} \hat{a}^{\dag}(\hat{a}^{\dag} \hat{a} + 1)^{n}\\
        &= \hat{a}^{\dag} \sum\limits_{n=0}^{\infty} (\frac{1}{\omega - (\hat{h}^{(0)} - \omega_{c} \hat{a}^{\dag}\hat{a} )})^{n+1}  (\omega_{c} \hat{a}^{\dag} \hat{a} + \omega_{c})^{n} \\
        &= \hat{a}^{\dag} \frac{1}{\omega - (\hat{h}^{(0)} - \omega_{c} \hat{a}^{\dag}\hat{a} ) - (\omega_{c} \hat{a}^{\dag} \hat{a} + \omega_{c})}.
    \end{split}
\end{equation}
In the second line we again shift the position of $n$ parentheses without moving any operator. In the third line we used the bosonic commutation relation $[\hat{a},\hat{a}^{\dag}]=1$, in the fourth line we used the commuting relation $[\hat{h}^{(0)} - \omega_{c} \hat{a}^{\dag}\hat{a}, ~ \hat{a}^{\dag}]=0$ to move $\hat{a}^{\dag}$ to the very left. In the fifth line we reverse the Dyson expansion, whose validity is guaranteed by the commutation relation $ [ \hat{h}^{(0)} , \omega_{c} \hat{a}^{\dag}\hat{a}  ] =0  $. Thus,we derive the following equation
\begin{equation}
    \hat{g}^{(0)}{(\omega)} \hat{a}^{\dag} = \hat{a}^{\dag} \hat{g}^{(0)}{(\omega - \omc)}. \label{eq.C3}
\end{equation}
Taking the Hermitian conjugate of both sides followed by a shift of argument $\omega \to \omega + \omc$, we find
\begin{equation}
    \hat{g}^{(0)}{(\omega)} \hat{a} = \hat{a} \hat{g}^{(0)}{(\omega + \omc)}. \label{eq.C4}
\end{equation}

\section{Resummation of interaction vertices}
\label{sec.resummation}
We first consider the vertex $\hat{G}^{(0)}{(E_\alpha +\omL)}\hat{D} \PEa$ in the term $\PEa \hat{D}^{\dag} \hat{G}^{(0)} \hat{D} \PEa$. It contains infinite orders of the interaction $\hat{U}$, and evaluates to
    \begin{equation}\label{k-space-GRPA-Stark-begin}
    \begin{split}
        &\hat{G}^{(0)}{(E_\alpha +\omL)}\hat{D}\PEa  \\
          &= g_{L}
          \sum_{{\bf q}, s}
          \sum\limits_{n=0}^{\infty} (\hat{g}^{(0)}_{(\Ea + \omL)} \hat{U})^n \hat{g}^{(0)}_{(\Ea + \omL)}
          \hat{b}_{{\bf q},s}^{\dag} \PEa \\
          &=g_{L}
          \sum_{{\bf q}, s}
          \sum\limits_{n=0}^{\infty} (\hat{g}^{(0)}_{(\Ea + \omL)} \hat{U})^n \hat{b}_{{\bf q},s}^{\dag}
          \hat{g}^{(0)}_{(\Ea - \Delta_{\bf q}^0)} \PEa \\
          &\approx g_{L} 
          \sum_{{\bf q}, s}
          \sum\limits_{n=0}^{\infty}   ~~
          \sum\limits_{ {\bf k}_1 , {\bf k}_2 , ... , {\bf k}_n } \hat{b}_{{\bf k}_n,s}^{\dag} 
          ( \hat{g}^{(0)}_{(\Ea - \Delta_{{\bf k}_n}^0) }   \hat{f}_{{\bf k}_n,{\bf k}_{n-1}}^{s}   )
          ( \hat{g}^{(0)}_{(\Ea - \Delta_{{\bf k}_{n-1}}^0)}   \hat{f}_{{\bf k}_{n-1},{\bf k}_{n-2}}^{s}   ) 
          ~ ... ~ \\
          & ~~~~~~~~~~~~~~~~~~~~ 
        ~ ... ~
          ( \hat{g}^{(0)}_{(\Ea - \Delta_{{\bf k}_2}^0)}   \hat{f}_{{\bf k}_2,{\bf k}_1}^{s}   )
          ( \hat{g}^{(0)}_{(\Ea - \Delta_{{\bf k}_1}^0)}   \hat{f}_{{\bf k}_1,{\bf q}}^{s}   )  
          ~
          \hat{g}^{(0)}_{(\Ea - \Delta_{\bf q}^0)} \PEa  \\
          &=  g_{L} 
          \sum_{{\bf q}, s}
          \sum\limits_{n=0}^{\infty}   ~~
          \sum\limits_{ {\bf k}_n } ~~ \hat{b}_{{\bf k}_n,s}^{\dag} ~
          [({\rm g}*{\rm f}^{s})^{n} * {\rm g}]_{ {\bf k}_n , {\bf q} } \PEa \\
          &=  g_{L} 
          \sum_{ {\bf q}' , s}
        ~~ \hat{b}_{{\bf q}',s}^{\dag} ~
          \sum_{{\bf q}} \sum\limits_{n=0}^{\infty}
          [({\rm g}*{\rm f}^{s})^{n} * {\rm g}]_{ {\bf q}' , {\bf q} } \PEa \\
          &= g_{L} 
          \sum_{{\bf q}', s} \hat{b}_{{\bf q}',s}^{\dag} ~ 
          \bigg( \sum_{\bf q}
          [  ({\rm g}^{-1} - {\rm f}^{s})^{-1} ]_{ {\bf q}' , {\bf q} }
          \bigg)  \PEa  
    \end{split}
\end{equation}
In the first line of Eq.~(\ref{k-space-GRPA-Stark-begin}), we insert the definition~(\ref{D}) of the driving operator $\hat{D}$, and then discard the de-excitation part of $\hat{D}$ because $\hat{b}_{{\bf q},s}\PEa=0$, which follows from the low-energy limit~(\ref{low-energy-approx}). 
We also expand $\hat{G}^{(0)}$ into a Born series containing infinite orders of $\hat{U}$, according to the Dyson expansion~(\ref{eq.series-expansion}). In the second line, we use Eq.~(\ref{g-commutator}) to move the far-right $\hat{b}^{\dag}$ leftward.
In the third line, we keep moving this $\hat{b}^{\dag}$ further leftward until it is adjacent to the $\hat{b}$ at the far-left. We use Eq.~(\ref{g-commutator}) whenever $\hat{b}^{\dag}$ crosses $\hat{g}^{(0)}$, and the approximation~(\ref{U-b-commutator-reduced})  whenever $\hat{b}^{\dag}$ crosses $\hat{U}$. 
For the $n$-th order expansion of $\hat{G}^{(0)}$, 
the repeated application of Eq.~(\ref{U-b-commutator-reduced}) creates summations over $n$ internal momenta (${\bf k}_1,{\bf k}_2, ... ,{\bf k}_n$), and the momentum carried by $\hat{b}^{\dag}$ is changed from the initial $\bf q$ to the final ${\bf k}_n$.
In the fourth line, we introduce the matrices ${\rm f}^s$ and ${\rm g}$, with operator-valued matrix elements
\footnote{Here, we define the multiplication between two matrices (with operator-valued matrix elements) $\rm A$ and $\rm B$ as
$$ [{\rm A} * {\rm B}]_{{\bf k'},{\bf k}} \equiv \sum\nolimits_{\bf k''} 
    ~ [{\rm A} ]_{{\bf k'},{\bf k''}}  ~ [{\rm B}]_{{\bf k''},{\bf k}} $$
where we sum over $\bf k''$ in the first Brillioun zone.}
\begin{equation}\label{operator-valued-matrix-element}
    \begin{split}
        [{\rm g}]_{ {\bf k} , {\bf q} } &\equiv \delta_{ {\bf k} , {\bf q} } \hat{g}^{(0)}_{(\Ea - \Delta_{\bf q}^0)}, \\
        [{\rm f}^{s}]_{ {\bf k} , {\bf q} } &\equiv \hat{f}_{{\bf k},{\bf q}}^{s}.
    \end{split}
\end{equation}
where $\hat{f}_{{\bf k},{\bf q}}^{s}$ is defined in Eq.~(\ref{eq.f-def}). Hence, the inverse matrix of $\rm g$ is simply
$
    [ {\rm g}^{-1} ]_{{\bf k},{\bf q}} = \delta_{{\bf k},{\bf q}} (\hat{g}^{(0)}_{(\Ea - \Delta_{\bf q}^0)})^{-1}.
$ 
We can directly check that this ${\rm g}^{-1}$ satisfies the definition of inverse, i.e., $[{\rm g} * {\rm g}^{-1}]_{{\bf k'},{\bf k}}=[{\rm g}^{-1} * {\rm g}]_{{\bf k'},{\bf k}}=\delta_{{\bf k'},{\bf k}}$. 
In the fifth line of Eq.~(\ref{k-space-GRPA-Stark-begin}), we replace the dummy index ${\bf k}_n$ by ${\bf q}'$, and rearrange the sequence of summation. 
In the sixth (final) line of Eq.~(\ref{k-space-GRPA-Stark-begin}), we use the formula
\begin{equation*}
    \sum\limits_{n=0}^{\infty}  ({\rm g} * {\rm f}^{s})^{n} * {\rm g} =    ({\rm g}^{-1} - {\rm f}^{s})^{-1} 
\end{equation*}
which is a matrix Taylor expansion (i.e. a matrix Born series). 

To further simplify the structure in the final line of Eq.~(\ref{k-space-GRPA-Stark-begin}), we recombine the matrix ${\rm f}^s$ and ${\rm g}$ by two new matrices ${\rm f}_{F}^s$ and ${\rm g}_{H}^{s}$, respectively defined as
\begin{equation}\label{split-gdelta-fF}
\begin{split}
    [{\rm f}_{F}^s]_{ {\bf k} , {\bf q} } 
    &= - \frac{U_{12}}{N} \hat{n}_{{\bf q},1,s} \equiv [{\rm f}_{F}^s]_{ * , {\bf q} } ~~ \text{independent of} ~ {\bf k} \\
    [{\rm g}_{H}^{s}]_{ {\bf k} , {\bf q} } 
    &=  \delta_{{\bf k},{\bf q}} 
    \bigg(   \big[\hat{G}^{(0)}_{(\Ea - \Delta_{\bf q}^0)} \big]^{-1}
    + U_{11}\hat{\nu}_{\Bar{s}}
    - U_{12}\sum\limits_{s'}\hat{\nu}_{s'}
    \bigg)^{-1}
\end{split}
\end{equation}
Here the character $F$ and $H$ stands for "Fock" and "Hartree". Note that the new matrix element $[{\rm f}_{F}^s]_{ {\bf k} , {\bf q} }$ depends only on its second momentum index $\bf q$, thus it can be represented by a simpler symbol, denoted by $[{\rm f}_{F}^s]_{ * , {\bf q} }$ in Eq.~(\ref{split-gdelta-fF}).

According to Eq.~(\ref{split-gdelta-fF}), the expression $\big( {\rm g}^{-1} - {\rm f}^{s} \big)$ in Eq.~(\ref{k-space-GRPA-Stark-begin}) can be recombined as $\big( ({\rm g}_{H}^{s})^{-1} - {\rm f}_{F}^s \big)$, because
\begin{equation*}
    \begin{split}
        [{\rm g}^{-1}]_{ {\bf k} , {\bf q} } - [({\rm g}_{H}^{s})^{-1}]_{ {\bf k} , {\bf q} } &= 
        \delta_{ {\bf k} , {\bf q} } \big( \hat{U} - U_{11}\hat{\nu}_{\Bar{s}} + U_{12}\sum\limits_{s'}\hat{\nu}_{s'} \big) \\
        &= [{\rm f}^{s}]_{ {\bf k} , {\bf q} }  -  [{\rm f}_{F}^s]_{ {\bf k} , {\bf q} } 
    \end{split}
\end{equation*}
where in the first line we use $(\hat{G}^{(0)})^{-1} = \hat{g}^{-1} - \hat{U}$, and the second line directly follows from the definition (\ref{eq.f-def}). 

This recombination (\ref{split-gdelta-fF}) allows the following evaluation of the term in parenthesis in the last line of Eq.~(\ref{k-space-GRPA-Stark-begin}),
\begin{widetext}
\begin{equation}\label{k-space-GRPA-collection}
    \begin{split}
        \sum_{\bf q} [  ({\rm g}^{-1} - {\rm f}^{s})^{-1} ]_{ {\bf q}' , {\bf q} } 
        &= \sum_{\bf q} [ \bigg( ({\rm g}_{H}^{s})^{-1} - {\rm f}_{F}^s \bigg)^{-1} ]_{ {\bf q}' , {\bf q} } \\
        &= \sum\limits_{\bf q} [{\rm g}_{H}^{s}]_{ {\bf q}' , {\bf q} } + \sum\limits_{\bf q} [{\rm g}_{H}^{s} * {\rm f}_{F}^s * {\rm g}_{H}^{s} ]_{ {\bf q}' , {\bf q} }
        + \sum\limits_{\bf q} [{\rm g}_{H}^{s} * {\rm f}_{F}^s * {\rm g}_{H}^{s}* {\rm f}_{F}^s * {\rm g}_{H}^{s}]_{ {\bf q}' , {\bf q} } + ~ ... \\
        & = ~~~   [{\rm g}_{H}^{s}]_{ {\bf q'} , {\bf q'} }  
        ~ + ~ [{\rm g}_{H}^{s}]_{ {\bf q'} , {\bf q'} } \sum\limits_{\bf q} [{\rm f}_{F}^s]_{ {\bf q'} , {\bf q} } [{\rm g}_{H}^{s}]_{ {\bf q} , {\bf q} } \\
        & ~~~   
        + [{\rm g}_{H}^{s}]_{ {\bf q'} , {\bf q'} }
        \sum\limits_{{\bf q}_1}
        [{\rm f}_{F}^s]_{ {\bf q'} , {\bf q}_1 }  [{\rm g}_{H}^{s}]_{ {\bf q}_1 , {\bf q}_1 }
        \sum\limits_{\bf q}
        [{\rm f}_{F}^s]_{ {\bf q}_1 , {\bf q} } [{\rm g}_{H}^{s}]_{ {\bf q} , {\bf q} } \\
        & ~~~   
        + [{\rm g}_{H}^{s}]_{ {\bf q'} , {\bf q'} }
        \sum\limits_{{\bf q}_1 }
        [{\rm f}_{F}^s]_{ {\bf q'} , {\bf q}_1 }  [{\rm g}_{H}^{s}]_{ {\bf q}_1 , {\bf q}_1 }
        \sum\limits_{ {\bf q}_2 }
        [{\rm f}_{F}^s]_{ {\bf q}_1 , {\bf q}_2 } [{\rm g}_{H}^{s}]_{ {\bf q}_2 , {\bf q}_2 }
        \sum\limits_{\bf q}
        [{\rm f}_{F}^s]_{ {\bf q}_2 , {\bf q} } [{\rm g}_{H}^{s}]_{ {\bf q} , {\bf q} }
        ~~~ + ...\\
        &=  [{\rm g}_{H}^{s}]_{ {\bf q'} , {\bf q'} } 
        ~~ + ~~
        [{\rm g}_{H}^{s}]_{ {\bf q'} , {\bf q'} } \bigg( \sum\limits_{\bf q} [{\rm f}_{F}^s]_{ * , {\bf q} } [{\rm g}_{H}^{s}]_{ {\bf q} , {\bf q} } \bigg) \\
        &~~~ +
        [{\rm g}_{H}^{s}]_{ {\bf q'} , {\bf q'} }
        \bigg( \sum\limits_{{\bf q}_1}
        [{\rm f}_{F}^s]_{ * , {\bf q}_1 }  [{\rm g}_{H}^{s}]_{ {\bf q}_1 , {\bf q}_1 } \bigg)
        \bigg( \sum\limits_{\bf q}
        [{\rm f}_{F}^s]_{ * , {\bf q} } [{\rm g}_{H}^{s}]_{ {\bf q} , {\bf q} } \bigg) \\
        &~~~ + 
        [{\rm g}_{H}^{s}]_{ {\bf q'} , {\bf q'} }
        \bigg( \sum\limits_{{\bf q}_1}
        [{\rm f}_{F}^s]_{ * , {\bf q}_1 }  [{\rm g}_{H}^{s}]_{ {\bf q}_1 , {\bf q}_1 }
        \bigg)
        \bigg( \sum\limits_{ {\bf q}_2 }
        [{\rm f}_{F}^s]_{ * , {\bf q}_2 } [{\rm g}_{H}^{s}]_{ {\bf q}_2 , {\bf q}_2 }\bigg) 
        \bigg( \sum\limits_{\bf q} [{\rm f}_{F}^s]_{ * , {\bf q} } [{\rm g}_{H}^{s}]_{ {\bf q} , {\bf q} } \bigg) 
        ~~~ + ...\\
        & =[{\rm g}_{H}^{s}]_{ {\bf q'} , {\bf q'} } \sum\limits_{n=0}^{\infty} \bigg( \sum\limits_{ {\bf q}'' }  [{\rm f}_{F}^s]_{ * , {\bf q}'' }  [{\rm g}_{H}^{s}]_{ {\bf q}'' , {\bf q}'' }  \bigg)^n \\
        & = [{\rm g}_{H}^{s}]_{ {\bf q'} , {\bf q'} }  \bigg( 1 - \sum\limits_{ {\bf q}'' }  [{\rm f}_{F}^s]_{ * , {\bf q}'' }  [{\rm g}_{H}^{s}]_{ {\bf q}'' , {\bf q}'' }  \bigg)^{-1}. 
    \end{split}
\end{equation}
\end{widetext}
In the first line we replace ${\rm g}^{-1} - {\rm f}^{s} $ by $ ({\rm g}_{H}^{s})^{-1} - {\rm f}_{F}^s $, in the second line we use the matrix Taylor expansion. In the third line we expand the matrix multiplication, and then use the diagonal property of matrix ${\rm g}_{H}^{s}$ to reduce the number of momentum indices to be summed. In the fourth line we use $[{\rm f}_{F}^s]_{ {\bf k} , {\bf q} } = [{\rm f}_{F}^s]_{ * , {\bf q} }$, then the summation over momentum indices decouple from one another, and can be evaluated separately (as we did in this line). All these parenthesized terms are identical, and thus we collect them order by order and in the sixth (final) line replace this infinite summation by the inverse of a single operator.

Based on Eqs.~(\ref{k-space-GRPA-collection}) and (\ref{k-space-GRPA-Stark-begin}), we obtain 
\begin{equation}\label{eq.k-space-GRPA-Stark-middle.appendix}
    \begin{split}
        &\PEa \hat{D}^{\dag}\hat{G}^{(0)}_{(E_\alpha +\omL)}\hat{D} \PEa \\
        &= \vert g_{L} \vert^2 \PEa 
          \sum_{{\bf q}'', s'} \hat{b}_{{\bf q}'',s'}
          \sum_{{\bf q}', s} \hat{b}_{{\bf q}',s}^{\dag} ~ 
          \bigg( \sum_{\bf q}
          [  ({\rm g}^{-1} - {\rm f}^{s})^{-1} ]_{ {\bf q}' , {\bf q} }
          \bigg)  \PEa \\
        &= \vert g_{L} \vert^2 \PEa 
         \sum_{{\bf q}', s} \hat{n}_{{\bf q}',1,s} ~ 
          \bigg( \sum_{\bf q}
          [  ({\rm g}^{-1} - {\rm f}^{s})^{-1} ]_{ {\bf q}' , {\bf q} }
          \bigg)  \PEa \\
        &= \vert g_{L} \vert^2 \PEa
          \sum_{{\bf q}', s} \hat{n}_{{\bf q'},1,s} ~
          [{\rm g}_{H}^{s}]_{ {\bf q'} , {\bf q'} } 
          \bigg( 1 - \sum\limits_{ {\bf q}'' }  [{\rm f}_{F}^s]_{ * , {\bf q}'' }  [{\rm g}_{H}^{s}]_{ {\bf q}'' , {\bf q}'' }  \bigg)^{-1}
            \PEa, 
    \end{split}
\end{equation}
where last line is Eq.~(\ref{eq.k-space-GRPA-Stark-middle}). 
In the first line we insert Eq.~(\ref{k-space-GRPA-Stark-begin}), as well as the definition of $\hat{D}^{\dag}$ and use the low-energy restriction $\PEa \hat{b}^{\dag} = 0$. In the second line we use $\PEa \hat{b}_{{\bf q}'',s'} \hat{b}_{{\bf q}',s}^{\dag} = \delta_{ {\bf q}'' , {\bf q}' } \delta_{ s' , s } \PEa \hat{c}_{{\bf q}',1,s}^{\dag} \hat{c}_{{\bf q}',1,s}$, which also follows from the low-energy restriction (\ref{low-energy-approx}). In the third (last) line we insert Eq.~(\ref{k-space-GRPA-collection}). 

\section{Mean field Hartree approximation and effective low-energy Hamiltonian}
\label{sec.MF}

\subsection{Hartree approximation}
To further simplify Eq.~(\ref{eq.k-space-GRPA-Stark-middle}), we next eliminate the operator $\hat{G}^{(0)}$ in $ [{\rm g}_{H}^{s}]_{ {\bf q} , {\bf q} } $ using Eq.~(\ref{G1-reduced-to-number}). 
We find
\begin{equation}\label{reduce-gHs-filling}
\begin{split}
    [{\rm g}_{H}^{s}]_{ {\bf q} , {\bf q} }  \PEa  
    &=  \bigg(   \big[\hat{G}^{(0)}_{(\Ea - \Delta_{\bf q}^0)} \big]^{-1}
    + U_{11}\hat{\nu}_{\Bar{s}}
    - U_{12}\sum\limits_{s'}\hat{\nu}_{s'}
    \bigg)^{-1} \PEa \\
    &= \sum_{n=0}^{\infty} \big(-U_{11}\hat{\nu}_{\Bar{s}}+U_{12}\sum\limits_{s'}\hat{\nu}_{s'} \big)^{n} \big[ \hat{G}^{(0)}_{(\Ea - \Delta_{\bf q}^0)} \big]^{n+1}  \PEa  \\
    &\approx \sum_{n=0}^{\infty} \big(-U_{11}\hat{\nu}_{\Bar{s}}+U_{12}\sum\limits_{s'}\hat{\nu}_{s'} \big)^{n} (\frac{-1}{\Delta_{\bf q}^0})^{n+1}  \PEa  \\
    &=\big(   -\Delta_{\bf q}^0
    + U_{11}\hat{\nu}_{\Bar{s}}
    - U_{12}\sum\limits_{s'}\hat{\nu}_{s'}
    \big)^{-1} \PEa \\
    &\approx \big(   -\Delta_{\bf q}^0
    + U_{11}\nu_{\Bar{s}}
    - U_{12}\sum\limits_{s'}\nu_{s'}
    \big)^{-1} \PEa \\
\end{split}
\end{equation}
where in the first line we insert into the definition of ${\rm g}_{H}^{s}$ in (\ref{split-gdelta-fF}). In the second line we apply a Dyson expansion of the operator, in this expansion, we can put all $\hat{G}^{(0)}$ to the far-right because $\hat{G}^{(0)}$ commutes with the electronic occupation in the lower band, $[\hat{G}^{(0)},\hat{\nu}_{s}]=0$. In the third line we use Eq.~(\ref{G1-reduced-to-number}) to reduce $\hat{G}^{(0)}$ into the denominator, which eliminates the degree of freedom of the upper band and the cavity. In the fourth line we turn the infinite Dyson expansion back to the inverse of a single operator. In the last line, we replace each filling operator $\hat{\nu}_s$ by its expectation value, which is the spin-resolved filling factor $\nu_s$. This approximation (\ref{mean-field-global}) treats filling factors as mean fields, which reduces $[{\rm g}_{H}^{s}]_{ {\bf q} , {\bf q} }$ to a screened denominator when it lies adjacent to $\PEa$.

Based on Eq.~(\ref{reduce-gHs-filling}), in Eq.~(\ref{eq.k-space-GRPA-Stark-middle.appendix}) we further simplify
\begin{equation}\label{reduce-gHs-single-occupation}
    \begin{split}
        &\sum\limits_{ {\bf q} }  [{\rm f}_{F}^s]_{ * , {\bf q} }  [{\rm g}_{H}^{s}]_{ {\bf q} , {\bf q} } \PEa \\ 
        &\approx - \frac{U_{12}}{N} \sum_{ {\bf q} } \hat{n}_{{\bf q},1,s}  \frac{ 1 } 
        {   -\Delta_{\bf q}^0 + U_{11}\nu_{\Bar{s}} - U_{12}\sum\limits_{s'}\nu_{s'}
        } \PEa  \\
        &\approx - \frac{U_{12}}{N} \sum_{ {\bf q} } \langle \hat{n}_{{\bf q},1,s} \rangle \frac{ 1 } {   -\Delta_{\bf q}^0 + U_{11}\nu_{\Bar{s}} - U_{12}\sum\limits_{s'}\nu_{s'}
        } \PEa  \\
    \end{split}
\end{equation}
where in the first line we insert the approximation (\ref{reduce-gHs-filling}). In the second line we further replace the number operator of the lower-band electron $\hat{n}_{{\bf q},1,s}$ by its expectation value. This is the second step of the mean-field decoupling~(\ref{mean-field-local}). Based on the approximation (\ref{reduce-gHs-single-occupation}) and (\ref{reduce-gHs-filling}), the expansion term~(\ref{eq.k-space-GRPA-Stark-middle.appendix}) in the effective Hamiltonian is finally reduced to
\begin{equation}\label{k-space-GRPA-Stark-final}
\begin{split}
        &\PEa \hat{D}^{\dag}\hat{G}^{(0)}_{(E_\alpha +\omL)}\hat{D} \PEa \\
        &\approx \vert g_{L} \vert^2 \PEa
        \frac{
        \sum_{{\bf q}', s} \hat{n}_{{\bf q'},1,s} ~
          \big(   -\Delta_{\bf q'}^0
    + U_{11}\nu_{\Bar{s}}
    - U_{12}\sum\limits_{s'}\nu_{s'}
    \big)^{-1}
        }{ 1 + \frac{U_{12}}{N} \sum_{ {\bf q''} } \langle \hat{n}_{{\bf q''},1,s} \rangle (   -\Delta_{\bf q''}^0 + U_{11}\nu_{\Bar{s}} - U_{12}\sum\limits_{s'}\nu_{s'}
        )^{-1} }
            \PEa  \\
        &\equiv - \vert g_{L} \vert^2 \PEa
        \sum_{{\bf q}, s}
        \frac{1}{\Delta_{{\bf q},s} }
        \hat{n}_{{\bf q},1,s} \PEa
\end{split}
\end{equation}
where in the last line, the screened denominator $\Delta_{{\bf q},s}$ is defined in Eq.~(\ref{renormalised-denominator}) of the main text.

\subsection{Low-energy Hamiltonian}
Above we have evaluated the term $\PEa \hat{D}^{\dag}\hat{G}^{(0)}\hat{D} \PEa$ in the Floquet Hamiltonian (\ref{eq.H_eff}), which results in a screened optical Stark shift. We next study the term $\PEa \hat{D}^{\dag}\hat{G}^{(0)}\hat{C}\hat{G}^{(0)}\hat{C}^{\dag}\hat{G}^{(0)}\hat{D} \PEa$ in Eq.~(\ref{eq.H_eff}). Based on the expression for the vertex
\begin{equation}\label{eq.vertex}
\hat{G}^{(0)}{(E_\alpha +\omL)}\hat{D}\PEa
\approx 
g_{L} \sum_{{\bf q}, s} \hat{b}_{{\bf q},s}^{\dag} ~ (-\Delta_{{\bf q},s} )^{-1}  \PEa
\end{equation}
which follows from Eqs.~(\ref{k-space-GRPA-Stark-begin}), (\ref{k-space-GRPA-collection}), (\ref{reduce-gHs-filling}) and (\ref{reduce-gHs-single-occupation}), we have
\begin{equation}\label{cav-med-int-final}
\begin{split}
&|g_c|^2 \PEa \hat{D}^\dagger \hat{G}^{(0)} _{(\Ea + \omL)} \hat{C}  \hat{G}^{(0)}_{(\Ea + \omL)} \hat{C}^\dagger \hat{G}^{(0)}_{(\Ea + \omL)} \hat{D} \PEa \\
&\approx 
\sum_{ \substack{{\bf q},s \\ {\bf q}',s' } } \frac{|g_L g_c|^2}{\Delta_{{\bf q},s} \Delta_{{\bf q}',s'}} 
\PEa \hat{b}_{{\bf q}',s'} 
\hat{C}  \hat{G}^{(0)}_{(\Ea + \omL)} \hat{C}^\dagger
\hat{b}_{{\bf q},s}^{\dag} \PEa \\
&= \sum_{ \substack{{\bf q},s, {\bf k},s'' \\ {\bf q}',s' ,{\bf k}',s''' } } \frac{|g_L g_c|^2}{N \Delta_{{\bf q},s} \Delta_{{\bf q}',s'}} 
\PEa \hat{b}_{{\bf q}',s'} 
\hat{a} \hat{b}^{\dag}_{{\bf k}',s'''}  \hat{G}^{(0)}_{(\Ea + \omL)} \hat{a}^\dagger \hat{b}_{{\bf k},s''}
\hat{b}_{{\bf q},s}^{\dag} \PEa \\
&= \sum_{ \substack{{\bf q},s \\ {\bf q}',s' } } \frac{|g_L g_c|^2}{N \Delta_{{\bf q},s} \Delta_{{\bf q}',s'}} 
\PEa \hat{c}_{{\bf q}',1,s'}^{\dag} 
\hat{c}_{{\bf q}',1,s'}  \hat{a} \hat{G}^{(0)}_{(\Ea + \omL)} \hat{a}^\dagger \hat{c}_{{\bf q},1,s}^{\dag}
\hat{c}_{{\bf q},1,s} \PEa \\
&= \sum_{ \substack{{\bf q},s \\ {\bf q}',s' } } \frac{|g_L g_c|^2}{N \Delta_{{\bf q},s} \Delta_{{\bf q}',s'}} 
\PEa \hat{c}_{{\bf q}',1,s'}^{\dag} 
\hat{c}_{{\bf q}',1,s'}  \hat{G}^{(0)}_{(\Ea - \Delta_c)} \hat{c}_{{\bf q},1,s}^{\dag}
\hat{c}_{{\bf q},1,s} \PEa \\
&\approx \sum_{ \substack{{\bf q},s \\ {\bf q}',s' } } \frac{|g_L g_c|^2}{N \Delta_{{\bf q},s} \Delta_{{\bf q}',s'}} 
\PEa \hat{c}_{{\bf q}',1,s'}^{\dag} 
\hat{c}_{{\bf q}',1,s'}  \hat{c}_{{\bf q},1,s}^{\dag}
\hat{c}_{{\bf q},1,s} \hat{G}^{(0)}_{(\Ea - \Delta_c)}  \PEa \\
&\approx - ~ \frac{1}{N}  \sum_{\substack{{\bf k}, s \\ {\bf k'}, s'}} \frac{\vert g_L g_c \vert^2 }{ \Delta_c  \Delta_{{\bf k'},s'}   \Delta_{{\bf k},s}  } \hat{c}_{{\bf k'},1, s'} ^{\dag}  \hat{c}_{{\bf k'},1, s'} \hat{c}_{{\bf k},1, s}^{\dag}  \hat{c}_{{\bf k},1, s}.
\end{split}
\end{equation}
In the first line we insert Eq.~(\ref{eq.vertex}), in the second line we use the definition of $\hat{C}$ in Eq.~(\ref{eq.C-def}), in the third line we use the low-energy restriction $ \hat{b}_{{\bf k},s''} \hat{b}_{{\bf q},s}^{\dag} \PEa = \delta_{ {\bf k} , {\bf q} } \delta_{ s'' , s } \hat{c}_{{\bf q},1,s}^{\dag} \hat{c}_{{\bf q},1,s} \PEa$, in the fourth line we use Eq.~(\ref{g-commutator}) to move $\hat{a}^{\dag}$ leftward, and then use the low-energy restriction $\PEa \hat{a} \hat{a}^{\dag} = \PEa$. In the fifth line we assume $[\hat{G}^{(0)}_{(\Ea - \Delta_c)},\hat{c}_{{\bf q},1, s}^{\dag}  \hat{c}_{{\bf q},1, s}]\approx0$, which ignores the screening effect on the cavity mode given by the electron on-site repulsion. In the sixth (last) line we use Eq.~(\ref{G1-reduced-to-number}) to replace $\hat{G}^{(0)}$ by the inverse of the laser-cavity detuning, which evaluates this Floquet Hamiltonian term into the cavity-mediated interaction.

We finally focus on the term $\PEa \hat{D}\hat{G}^{(0)}_{(\Ea-\omL)}\hat{D}^{\dag} \PEa$ in the Floquet Hamiltonian (\ref{eq.H_eff}). Following the same procedure in the previous evaluations, we find this term results in another energy-shift, with the same form as Eq.~(\ref{k-space-GRPA-Stark-final}), apart from a substitution $\Delta_{\bf q}^0 \to \Delta_{\bf q}^0 + 2 \omL$. Specifically, this means
\begin{equation}\label{BS-final}
\PEa \hat{D}\hat{G}^{(0)}_{(E_\alpha -\omL)}\hat{D}^{\dag} \PEa 
\approx 
- \vert g_{L} \vert^2 \PEa \sum_{{\bf q}, s}
\frac{1}{\Delta_{{\bf q},s}^{BS} }
\hat{n}_{{\bf q},1,s} \PEa
\end{equation}
which is the screened Bloch-Siegert shift, with denominator $\Delta_{{\bf q},s}^{BS}$ defined in Eq.~(\ref{eq.BS-denominator}).

Inserting the result of Eqs.~(\ref{k-space-GRPA-Stark-final}), (\ref{cav-med-int-final}) and (\ref{BS-final}) into Eq.~(\ref{eq.H_eff}), we construct the low-energy Floquet Hamiltonian in Eq.~(\ref{eq.main-result}). Note that the $\hat{U}_{11}$ term in Eq.~(\ref{eq.U_eff}) comes from the direct projection term $\PEa \hat{H} \PEa$ in Eq.~(\ref{eq.H_eff}).

\subsection{Corrections to the low-energy Hamiltonian}\label{appendix-Corrections}
Below we estimate the corrections to the Floquet Hamiltonian~(\ref{eq.main-result}). 
First, we analyse the accuracy of Eq.~(\ref{G1-reduced-to-number}) as the electron-cavity coupling becomes ultra-strong and cavity-induced changes to the ground state manifold can no longer be ignored. Since $[\hat{G}^{(0)}]^{-1} = [\hat{G}]^{-1} + i g_c (\hat{C} - \hat{C}^{\dag}) $, we have
\begin{equation}\label{G1-reduce-explain}
\begin{split}
\hat{G}^{(0)}_{( \Ea - \Delta^{0}_{\bf k})}  \PEa
& = \bigg(1 + i g_c \hat{G}^{(0)}_{( \Ea - \Delta^{0}_{\bf k})} (\hat{C}-\hat{C}^{\dag})  \bigg) \hat{G}_{( \Ea - \Delta^{0}_{\bf k})}  \PEa  \\
& \approx - ( \Delta^{0}_{\bf k})^{-1} \PEa 
- i g_c ( \Delta^{0}_{\bf k})^{-1}  \hat{G}^{(0)}_{( \Ea - \Delta^{0}_{\bf k})} (\hat{C}-\hat{C}^{\dag})  \PEa
\end{split}
\end{equation}
In the second line we use the definition of the projector $\PEa = \delta(\Ea - \hat{H})$ to replace $\hat{G}_{( \Ea - \Delta^{0}_{\bf k})} $ by $- ( \Delta^{0}_{\bf k})^{-1}$. Under weak electron-cavity coupling, Eq.~(\ref{low-energy-approx}) implies $\hat{C}\PEa=\hat{C}^{\dag}\PEa=0$, which reduces Eq.~(\ref{G1-reduce-explain}) to Eq.~(\ref{G1-reduced-to-number}). However, for ultra-strong electron-cavity coupling $g_c \sim \omL$, the electron-cavity vertex $\hat{C}$ in Eq.~(\ref{eq.C-def}) must include the counter-rotating terms (simultaneously creating cavity photon and band-excitation), and $\PEa$ will no longer project to the 0-excitation manifold as in Eq.~(\ref{low-energy-approx}). Consequently, the term $\hat{G}^{(0)}_{( \Ea - \Delta^{0}_{\bf k})} (\hat{C}-\hat{C}^{\dag})  \PEa$ in Eq.~(\ref{G1-reduce-explain}) will contribute to a correction with prefactor $1 /(\omc+\omL)$.

Next, as the cavity-electron term $\hat{C}$ is expanded only to the lowest order in Eq.~(\ref{eq.H_eff}),  terms of order $\vert g_c \vert^4$ such as $ \hat{ \cal{P} } \hat{D}^{\dag} \hat{C} \hat{G}^{(0)} \hat{C}^{\dag} \hat{G}^{(0)} \hat{C} \hat{G}^{(0)} \hat{C}^{\dag} \hat{G}^{(0)} \hat{D} \hat{ \cal{P} } $, as well as higher-order terms, are ignored in the effective Hamiltonian. 
These terms renormalize the laser-cavity detuning $\frac{1}{\Delta_c} \to \frac{1}{\Delta_c} (1 - \frac{g_c^2}{\Delta_c} \chi ) $ where $\chi$ denotes the Lindhard function for the inter-band response. They are thus analogous to inter-band polarization bubbles which appear in Feynman treatments. 
This screening is weak because the bubble comprises two electron-cavity vertices connected by an inter-band electron-hole propagator, (together with an additional photon propagator) giving rise to a factor $\sim \vert g_c \vert^2 / (\Delta_c \Delta_{{\bf q },s})$ which is negligible under the off-resonating driving condition considered in this work. 

We finally focus on the screening of the cavity-electron vertex by the electron repulsion, which we ignored in the fifth line of Eq.~(\ref{cav-med-int-final}) by assuming $[\hat{G}^{(0)}_{(\Ea - \Delta_c)},\hat{n}_{{\bf q},1, s} ]\approx0$. Since $\hat{U}_{12}$ and $\hat{U}_{22}$ has no influence in $\PEa \hat{n}_{{\bf q}',1, s'} \hat{G}^{(0)} \hat{n}_{{\bf q},1, s} \PEa $, we only need to evaluate 
\begin{equation}
\begin{split}
[\hat{U}_{11}, \hat{n}_{{\bf q} 1 s} ]
&= \frac{U_{11}}{N} \sum\limits_{{\bf k'},{\bf q'}} \hat{c}_{{\bf k'}-{\bf q'} 1 \Bar{s}}^{\dag}  \hat{c}_{{\bf k'} 1 \Bar{s}}
( \hat{c}_{{\bf q}+{\bf q'} 1 s}^{\dag} \hat{c}_{{\bf q} 1 s}
  -
\hat{c}_{{\bf q} 1 s}^{\dag} \hat{c}_{{\bf q}-{\bf q'} 1 s}) \\
&\approx \frac{U_{11}}{N} \sum\limits_{{\bf k'},({\bf q'}={\bf 0})} \hat{c}_{{\bf k'} 1 \Bar{s}}^{\dag}  \hat{c}_{{\bf k'} 1 \Bar{s}}
( \hat{c}_{{\bf q} 1 s}^{\dag} \hat{c}_{{\bf q} 1 s}
  -
\hat{c}_{{\bf q} 1 s}^{\dag} \hat{c}_{{\bf q} 1 s}) \\
&= 0 \\
\end{split}
\end{equation}
In the second line, consistent with our treatment in Eq.~(\ref{U-b-commutator-appendix}), we only retain the ${\bf q'}={\bf 0}$ part, which makes the expectation values of charge waves $\langle \hat{c}_{{\bf k'}-{\bf q'} 1 \Bar{s}}^{\dag}  \hat{c}_{{\bf k'} 1 \Bar{s}} \rangle$ and $\langle \hat{c}_{{\bf q}+{\bf q}' 1 s}^{\dag} \hat{c}_{{\bf q} 1 s} \rangle$ non-zero. Consequently, after cancelling the equivalent terms in the second line, we have $[\hat{U}_{11}, \hat{n}_{{\bf q} 1 s} ]\approx0$. Since $[\hat{g}^{(0)}_{(\Ea - \Delta_c)},\hat{n}_{{\bf q},1, s} ]=0$, we conclude from Eq.~(\ref{eq.Dyson-eq}) that, in the absence of the charge order in the lower-band, we can safely ignore the corrections given by the commutator $[\hat{G}^{(0)}_{(\Ea - \Delta_c)},\hat{n}_{{\bf q},1, s} ]$.

\subsection{GRPA calculation}
\label{sec.GRPA}

Below we compare our screened Floquet Hamiltonian~(\ref{eq.main-result}) with an alternative method using the Matsubara formalism, where we construct Feynman diagrams in the laser-rotating frame. We move to the laser-rotating frame by applying the following unitary transformation to the full driven Hamiltonian (\ref{eq.H_tot}), $H(t) \to U_{t} H(t) U_{t}^{\dag} + i\hbar (\partial_{t} U_{t})U^{\dag}_{t}$, and $\vert\psi\rangle_{t} \to U_{t}\vert\psi\rangle_{t}$, where
\begin{equation} \label{eq.U_t}
    U_{t}=e^{i \omega_{L} t (\hat{a}^{\dag}\hat{a} + \sum\limits_{\mathbf{q}, s} \hat{c}_{\mathbf{q} 2 s}^{\dag} \hat{c}_{\mathbf{q} 2 s} )}.
\end{equation}
Then, after discarding the counter-rotating terms (i.e applying the RWA in the laser-matter interaction), the dipolar Hamiltonian becomes static in the rotating frame, 
\begin{equation}\label{H_Dip_r}
\begin{split}
\hat{H}_{\text {dip}}^{\text{rot}} &= \sum_{\mathbf{q}, s} \big( \epsilon_{\mathbf{q}, 1} -\mu \big) \hat{n}_{\mathbf{q} 1 s} + \big( \epsilon_{\mathbf{q}, 2} -\mu - \omega_L \big) \hat{n}_{\mathbf{q} 2 s} + (\omega_c-\omega_L) ~ \hat{a}^{\dag}\hat{a} \\
& +  \sum_{\mathbf{q}, s}  -i ~ (g_{L} -  g_{c} \hat{a} ) \hat{c}_{\mathbf{q} 2 s}^{\dag} \hat{c}_{\mathbf{q} 1 s}+\text { h.c. }   \\
& + \hat{U}_{11} + \hat{U}_{22} + \hat{U}_{12}.
\end{split}
\end{equation}
    \begin{figure}[t]
        \includegraphics[width=0.4\textwidth]{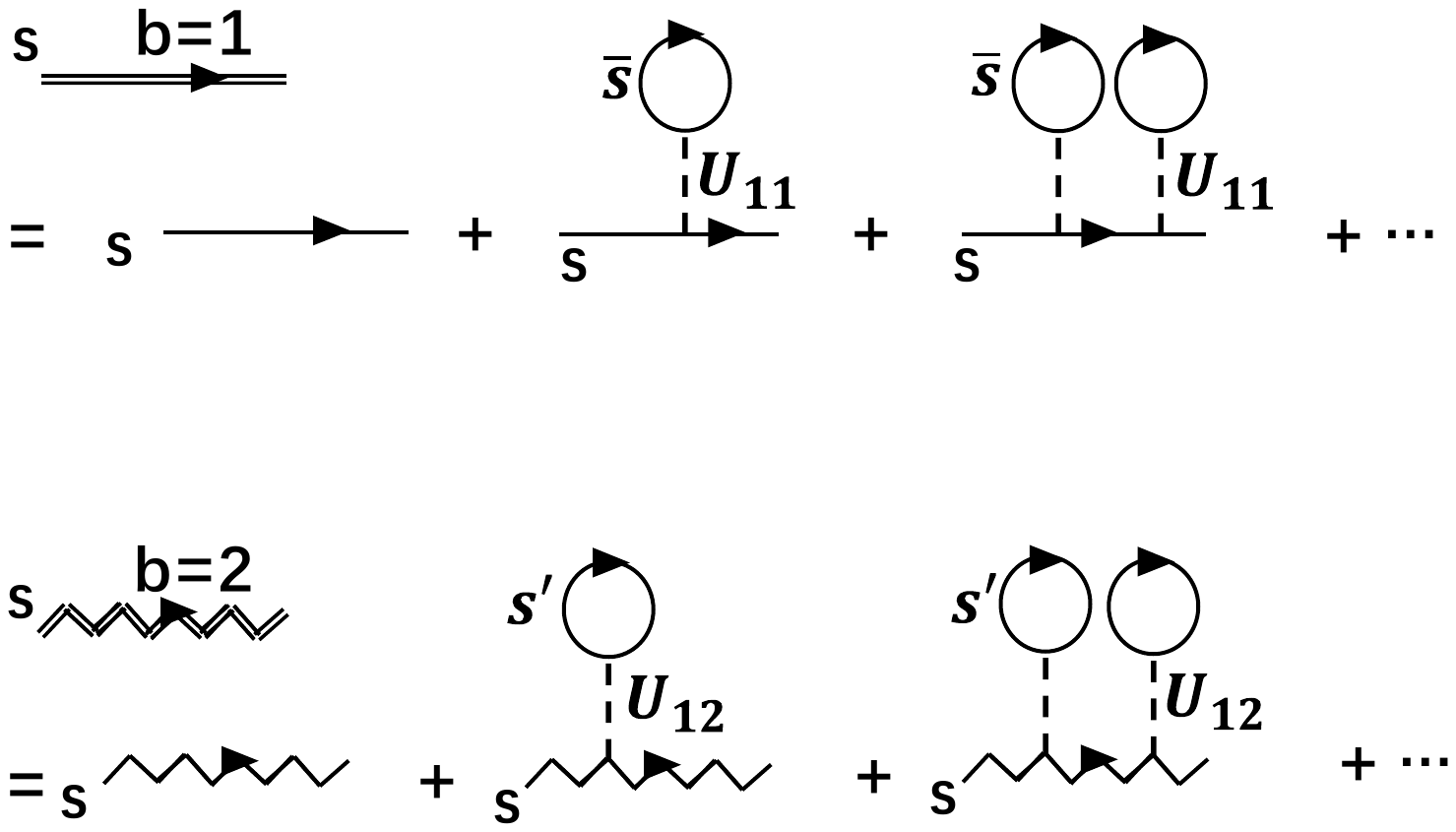}
        \captionsetup{justification=raggedright,singlelinecheck=false}
        \caption{The dressed single-particle propagators of two bands in the rotating frame. Note that the b=1 line will be used as the \textit{hole} propagator in the lower-band. The Hartree contribution of $U_{22}$ term is ignored in the b=2 line, for the reason discussed below Eq.~(\ref{U-b-commutator-reduced}).}
        \label{fig:single-particle-propagator}
    \end{figure}
   \begin{figure}[t]
        \includegraphics[width=0.4\textwidth]{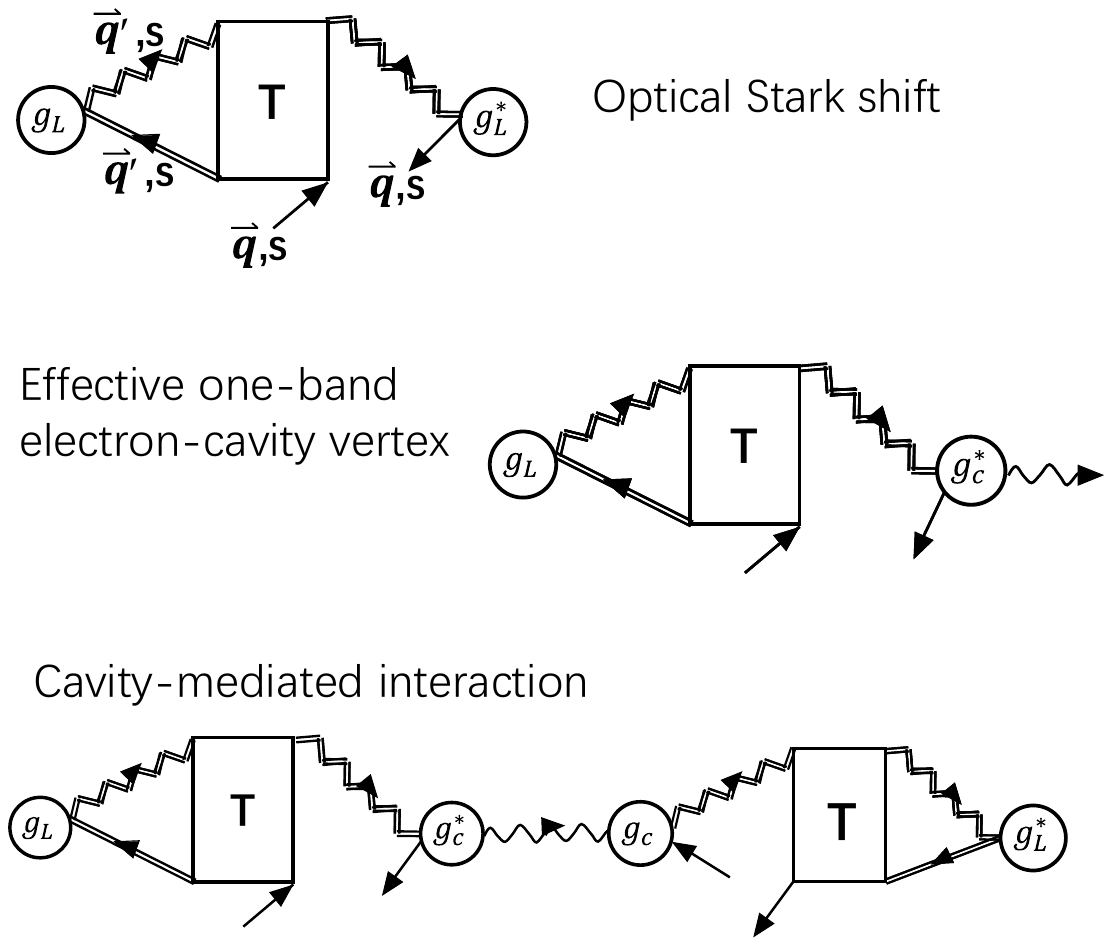}
        \captionsetup{justification=raggedright,singlelinecheck=false}
        \caption{The renormalised optical Stark shift, effective one-band electron-cavity vertex, and cavity-mediated interaction in the rotating frame. The wavy arrow denotes the cavity photon. The incoming and outgoing straight arrows denote the lower-band electron.
        }
        \label{fig:renormalisation-by-T}
    \end{figure}
We next apply the usual Feynman diagrammatic approach to this Hamiltonian to study its effective low-energy response. 
We will sum over a series of Feynman diagrams in the irreducible two-particle vertex, which in the static limit gives the equivalent result to Eq.~(\ref{renormalised-denominator}). As shown in Fig.~\ref{fig:single-particle-propagator}, the electron and hole propagators in the Hamiltonian~(\ref{H_Dip_r}) are first dressed by the Hartree terms. The dressed propagators for the lower- and upper-band respectively read
\begin{equation}
    \begin{split}
        \mathcal{G}_{1,{\bf q},s,i q_n} &= \frac{1}{i q_n - (\epsilon_{{\bf q},1} -\mu + U_{11} \nu_{1,\Bar{s}}) } \\
        \mathcal{G}_{2,{\bf q},s,i q_n} &= \frac{1}{i q_n - (\epsilon_{{\bf q},2} - \omega_L -\mu + U_{12} \sum\limits_{s'}\nu_{1,s'}) }
    \end{split}
\end{equation}
where the fermionic Matsubara Frequency is defined as $q_n = \frac{(2n+1)\pi}{\beta}$ for all integer $n \in (-\infty,\infty)$. At low temperature we assume $\beta\to\infty$. The asymmetry between $\mathcal{G}_{1}$ and $\mathcal{G}_{2}$ arises from the fact that the upper-band is empty, so that $\nu_{2,s}=0$. The Fock self-energy disappears in these single particle propagators, because we only consider on-site electron-electron interactions.

\begin{figure}
        \includegraphics[width=0.4\textwidth]{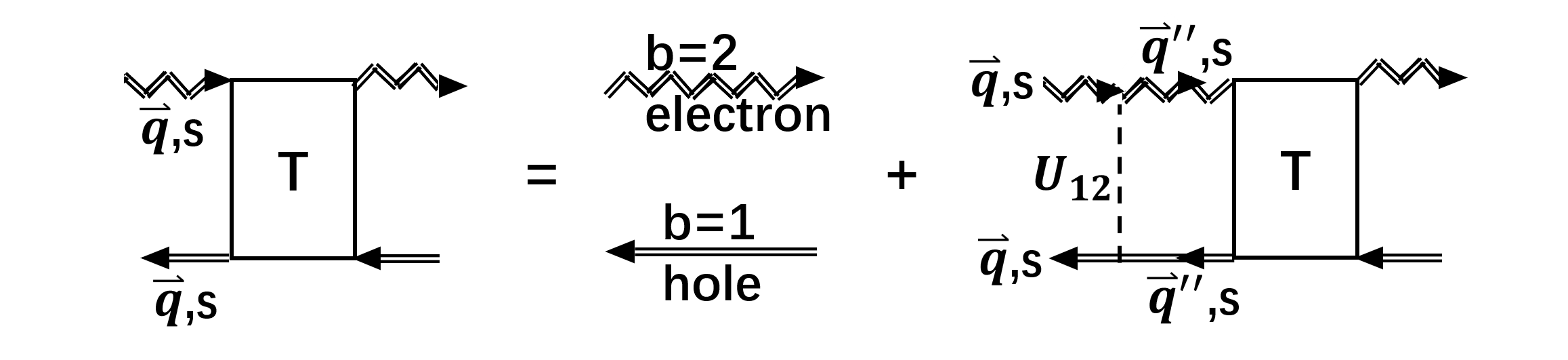}
        \captionsetup{justification=raggedright,singlelinecheck=false}
        \caption{The electron-hole t-matrix, which describes the attraction between the upper-band electron and the lower-band hole. The Matsubara frequency is not shown for simplicity.}
        \label{fig:particle-hole-transfer-matrix}
    \end{figure}

The effective attraction between the screened electrons and holes, represented by the so-called GRPA polarisation diagrams \cite{PhysRevB.40.3802}, are shown in Fig.~\ref{fig:renormalisation-by-T}. All of these diagrams contain an opened polarization bubble~\cite{PhysRevB.80.174401} with an inter-band electron-hole t-matrix~\cite{zimmermann1985mass}, as shown in Fig.~\ref{fig:particle-hole-transfer-matrix}. It reads
\begin{widetext}
\begin{equation}
    \begin{split}
        T &= \sum\limits_{n=0}^{\infty} \bigg( \frac{U_{12}}{\beta N} \sum\limits_{{\bf q''},i q_n} \frac{1}{i q_n - (\epsilon_{{\bf q''},1} -\mu + U_{11} \nu_{1,\Bar{s}}) } \frac{1}{i q_n - (\epsilon_{{\bf q''},2} - \omega_L -\mu + U_{12} \sum\limits_{s'}\nu_{1,s'}) } \bigg)^n \\
        &= \sum\limits_{n=0}^{\infty} \bigg( \frac{U_{12}}{ N} \sum\limits_{{\bf q''} } \frac{- n_F( \epsilon_{{\bf q''},1} -\mu + U_{11} \nu_{1,\Bar{s}} ) + n_F(\epsilon_{{\bf q''},2} - \omega_L -\mu) }{ (\epsilon_{{\bf q''},1} -\mu + U_{11} \nu_{1,\Bar{s}}) - (\epsilon_{{\bf q''},2} - \omega_L -\mu + U_{12} \sum\limits_{s'}\nu_{1,s'} )  } \bigg)^n \\
        &= \sum\limits_{n=0}^{\infty} \bigg( -\frac{U_{12}}{ N} \sum\limits_{{\bf q''} } \frac{ \langle \hat{n}_{{\bf q}'',1,s} \rangle }{ - \Delta_{\bf q''}^0 + U_{11} \nu_{1,\Bar{s}}
        - U_{12} \sum\limits_{s'}\nu_{1,s'}  } \bigg)^n \\
        &= \bigg(1 + \frac{U_{12}}{ N} \sum\limits_{{\bf q''} } \frac{ \langle \hat{n}_{{\bf q}'',1,s} \rangle }{ - \Delta_{\bf q''}^0 + U_{11} \nu_{1,\Bar{s}}
        - U_{12} \sum\limits_{s'}\nu_{1,s'}  } \bigg)^{-1} \\
    \end{split}
\end{equation}
\end{widetext}
where the Fermi distribution function is defined as $n_F(x) \equiv 1/(1+\exp(\beta x))$. In the third line, a geometric series similar to the infinite sum in Eq.~(\ref{k-space-GRPA-collection}) appears. Note that in the non-interacting model where $U_{12}\to0$, this electron-hole t-matrix reduces to unity $T\to1$, which is understood as a two-particle delta function (for both the upper-band electron line and the lower-band hole line). Meanwhile, note that when the exciton resonance requirement (\ref{exciton-resonance}) is fulfilled, it diverges, $T\to\infty$.

The $t$-matrix allows us to calculate the renormalised optical Stark shift, the effective one-band electron-cavity vertex, and the cavity-mediated interaction (but not the Bloch-Siegert shift). The expressions are represented by the Feynman diagrams in Fig.~\ref{fig:renormalisation-by-T}.
For example, the closed GRPA bubble for the optical Stark shift in Fig.~\ref{fig:renormalisation-by-T}, contributes to a self-energy term for the lower-band electron, whose value at the 4-momentum ${\bf q},i q_n$ evaluates to 
\begin{equation}
    \begin{split}
        \Sigma_{{\bf q}, iq_n} ~ = ~ \vert g_L \vert^2  \frac{1}{i q_n - (\epsilon_{{\bf q},2} - \omega_L -\mu + U_{12} \sum\limits_{s'}\nu_{1,s'}) } T ,
    \end{split}
\end{equation}
and thus the laser-dressed propagator of the lower-band electron reads
\begin{equation}
    \begin{split}
        & \mathcal{G}_{1,{\bf q},s,i q_n}^{\text{Fl}} = \frac{1}{\mathcal{G}_{1,{\bf q},s,i q_n}^{-1} - \Sigma_{{\bf q}, iq_n} }  \\
        &=    \frac{1}{i q_n - (\epsilon_{{\bf q},1} -\mu + U_{11} \nu_{1,\Bar{s}}  +   \frac{ \vert g_L \vert^2 T}{i q_n - (\epsilon_{{\bf q},2} - \omega_L -\mu + U_{12} \sum\limits_{s'}\nu_{1,s'}) }   ) }   
    \end{split}
\end{equation}
which has a pole at
\begin{widetext}
\begin{equation}
    \begin{split}
        i q_n & \to \epsilon_{{\bf q},1} -\mu + U_{11} \nu_{1,\Bar{s}}  +   \frac{ \vert g_L \vert^2 T}{i q_n - (\epsilon_{{\bf q},2} - \omega_L -\mu + U_{12} \sum\limits_{s'}\nu_{1,s'}) }     \\
        &\approx  \epsilon_{{\bf q},1} -\mu + U_{11} \nu_{1,\Bar{s}}  +   \frac{ \vert g_L \vert^2 T}{(\epsilon_{{\bf q},1} -\mu + U_{11} \nu_{1,\Bar{s}}) - (\epsilon_{{\bf q},2} - \omega_L -\mu + U_{12} \sum\limits_{s'}\nu_{1,s'}) } \\
        &=  \epsilon_{{\bf q},1} -\mu + U_{11} \nu_{1,\Bar{s}}  
        - \frac{ \vert g_L \vert^2 }
        { \big(\Delta_{\bf q}^0  - U_{11} \nu_{1,\Bar{s}}  + U_{12} \sum\limits_{s'}\nu_{1,s'}\big)
        \bigg(1 + \frac{U_{12}}{ N} \sum\limits_{{\bf q''} } \frac{ \langle \hat{n}_{{\bf q}'',1,s} \rangle }{ - \Delta_{\bf q''}^0 + U_{11} \nu_{1,\Bar{s}}
        - U_{12} \sum\limits_{s'}\nu_{1,s'}  } \bigg) }  \\
        &=  \epsilon_{{\bf q},1} -\mu + U_{11} \nu_{1,\Bar{s}}  - \frac{ \vert g_L \vert^2 }{ \Delta_{{\bf q},s}  }
    \end{split}
\end{equation}
\end{widetext}
which exactly reveals the lower-band's energy shift given by the screened optical Stark effect at spin-orbital $({\bf q},s)$ in Eq.~(\ref{eq.h_eff}). Compared with the optical Stark shift in the non-interacting case, an excitonic enhancement with factor $T$ is observed.
This shows that the renormalised denominator $\Delta_{{\bf q },s}$, which appears in the renormalized optical Stark shift and the cavity-mediated interaction in the effective Floquet Hamiltonian~(\ref{eq.main-result}), corresponds to the GRPA graphs in the retarded interaction formalism. 

\begin{figure}
        \includegraphics[width=0.45\textwidth]{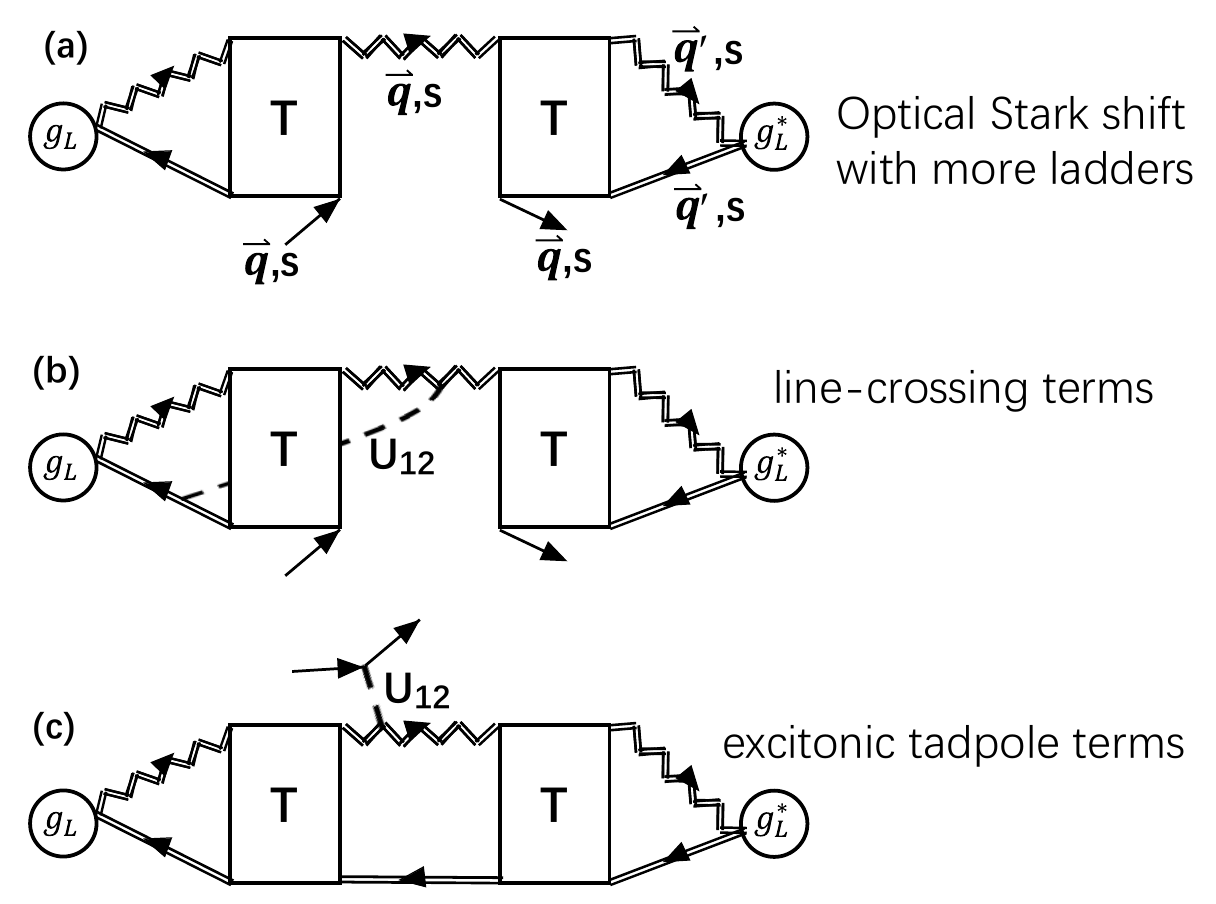}
        \captionsetup{justification=raggedright,singlelinecheck=false}
        \caption{The additional Ladder diagrams for the self-energy in the lower-band, which are not included in our result. Merely including the analytically tractable diagrams in (a) will give non-physical result, $(\Delta_{{\bf q},s})^{-1} \propto T^2$.}
        \label{fig:more-Ladder-graphs}
    \end{figure}

One can include more ladder diagrams in the self-energy calculation, e.g. as shown in Fig.\ref{fig:more-Ladder-graphs}(a), however, this would end up in a $T^2$ enhancement on the optical Stark shift $\Delta_{{\bf q},s}$ showing several nonphysical features: this enhancement is not consistent with the exact result in the flat-bandgap limit (see the right 3 panels in Fig.~\ref{fig.denominator}), and also, it restricts the sign of the detuning $\Delta_{{\bf q},s}$ to be non-negative even when $\omL > \omega_{ex}$. As discussed in Ref.~\cite{PhysRevB.40.3788}, in order to include all possible ladder diagrams, one must simultaneously include other diagrams, e.g. the "line-crossing" graphs in Fig.\ref{fig:more-Ladder-graphs}(b) and the "excitonic tadpole" graphs in Fig.\ref{fig:more-Ladder-graphs}(c), which could cancel the nonphysical effects, but are analytically uneasy to tract. In our Green operator approach, the $T^2$ enhancement can be reproduced by a consecutive mean-field decoupling on Eq.~(\ref{eq.k-space-GRPA-Stark-middle.appendix}), but before including these additional terms, we must first retain more terms in the commutator relation Eq.~(\ref{U-b-commutator-appendix}), so as to keep the resulting Floquet Hamiltonian valid in the flat-bandgap limit. This improvement is left for future study.

\subsection{ Comparison to Semiconductor Bloch equation}\label{Appendix:SBE-comparison}
Here we compare our Floquet method with the Semiconductor Bloch equation (SBE) method. The starting point for SBE is the RWA Hamiltonian (\ref{H_Dip_r}), from which the Heisenberg equations of motion (EOM) of the polarisation operators $\hat{b}_{{\bf k},s}$ and the occupancy operators $\hat{n}_{{\bf k},b,s}$ are obtained, i.e. $i\partial_t \hat{b}_{{\bf k},s} = [\hat{b}_{{\bf k},s},\hat{H}_{\text {dip}}^{\text{rot}}]$ and $i\partial_t \hat{n}_{{\bf k},b,s} = [\hat{n}_{{\bf k},b,s},\hat{H}_{\text {dip}}^{\text{rot}}]$. 
We further simplify the $[\hat{b}_{{\bf k},s},\hat{U}]$ commutators in these EOMs by the commutator approximation~(\ref{U-b-commutator-reduced}), and simplify the $[\hat{n}_{{\bf k},b,s},\hat{U}]$ commutators by the equivalent RPA-type treatment (see \cite{haug2009quantum}), which amounts to ignoring the scattering term in SBE.

Upon mean-field decoupling of these simplified EOMs, we will get the SBE (see, e.g. Ref.~\cite{haug2009quantum}). But once this mean-field approximation is made, we can no longer find the effective low-energy Hamiltonian from the resulting SBEs (i.e. mean-field differential equations). Since we are in the large-detuning limit, one may instead try if it is possible to skip the mean-field treatment in SBE, and adiabatically eliminate the original operator EOM for $\hat{b}_{{\bf k},s}$ to get the effective Hamiltonian: Specifically this means we take $i\partial_t \hat{b}_{{\bf k},s}\to0$, and then based on the commutator approximation (\ref{U-b-commutator-reduced}), we get an relation between $\hat{b}_{{\bf k},s}$ and $\hat{n}_{{\bf k},b,s}$, i.e.,
\footnote{For a clear comparison, below we ignore the cavity and only consider the laser-material coupling, and we also assume the hole-doping in the lower-band is very weak.}
\begin{equation*}
\sum\limits_{\bf k'} 
\big[\Gamma_{\text{ACS}} \big]_{{\bf k},{\bf k'}} \hat{b}_{{\bf k'},s}^{\dag} -i g_L^* ( \hat{n}_{{\bf k},2,s} - \hat{n}_{{\bf k},1,s} ) \to 0
\end{equation*}
where $\Gamma_{\text{ACS}}$ is a matrix with elements defined as
\begin{equation*}
\big[\Gamma_{\text{ACS}} \big]_{{\bf k},{\bf k'}} = (\ep_{{\bf k'},21}-\omL-U_{11}+2U_{12})\delta_{{\bf k},{\bf k'}} - \frac{U_{12}}{N}.
\end{equation*}
This relation results in
\begin{equation} \label{eq.replacement}
\hat{b}_{{\bf k},s}^{\dag} \to i g_L^* \sum\limits_{\bf k'} \left[\frac{1}{\Gamma_{\text{ACS}}} \right]_{{\bf k},{\bf k'}} ( \hat{n}_{{\bf k},2,s} - \hat{n}_{{\bf k},1,s} )
\end{equation}
which allows us to replace the operator $\hat{b}_{{\bf k},s}$ and $\hat{b}_{{\bf k},s}^{\dag}$ in Hamiltonian (\ref{H_Dip_r}) by $\hat{n}_{{\bf k},b,s}$.
\footnote{The replacement~(\ref{eq.replacement}) in Eq.~(\ref{H_Dip_r}) gives exactly the same excitonically enhanced Stark shift as our Floquet Hamiltonian result in Eq.~(\ref{k-space-GRPA-Stark-final}). However, note that $\hat{b}_{{\bf k},s}$ is also contained in Eq.~(\ref{H_Dip_r}), including the substitution $\hat{b}_{{\bf k},s} \to ...$ leads to the error compared with our Floquet Hamiltonian result.} 
The effective Hamiltonian $\hat{H}'$ given by this adiabatic elimination produces the same excitonically enhanced optical Stark shift, but it is too large by a factor 2 compared with our Floquet Hamiltonian result. The reason for this error is clear: after the elimination of $\hat{b}_{{\bf k},s}$, there is still upper-band degree of freedom in $\hat{H}'$, i.e. the term $ (\ep_{{\bf q}, 2} -\mu - \omega_L) \hat{n}_{{\bf q},2,s}$, which is not eliminated. This error can in principle be fixed, if we instead eliminate the EOM of the operator $\hat{c}_{{\bf k},2,s}$ rather than the EOM of $\hat{b}_{{\bf k},s}$, such that the upper-band degree of freedom is completely eliminated. 

In short, the mean field SBE cannot be used to derive the effective low-energy Hamiltonian derived in this manuscript, but the adiabatic elimination of operator EOM can, in principle, achieve this. 


\section{Low-temperature absorbance}\label{sec.absorbance}
The optical absorbance spectrum $\alpha(\omega)$ for our model is defined as the imaginary part of the dipole-dipole correlation function in frequency domain, which, in the low-temperature limit, reads~\cite{matsueda2005excitonic,PhysRevB.29.4401}
\begin{equation}\label{eq.absorbance}
\begin{split}
\alpha(\omega) &= -\frac{1}{\pi} \text{Im} \langle G \vert \frac{1}{g_L^*} \hat{D}^{\dag} \frac{1}{\omega+E_G-\hat{H} + i \gamma } \frac{1}{g_L} \hat{D} \vert G \rangle \\
&= - \frac{1}{\vert g_L \vert^{2} \pi} \text{Im} \langle G \vert \hat{\mathcal{P}}_{E_G} \hat{D}^{\dag} \hat{G}_{(\omega+E_G + i \gamma)} \hat{D} \hat{\mathcal{P}}_{E_G} \vert G \rangle. 
\end{split}
\end{equation}
Here $\gamma$ is a tiny positive number broadening the absorbance spectrum, $\vert G \rangle$ is the ground state of the static Hamiltonian $\hat{H}$ with eigenenergy $E_G$, and in the second line we use $\hat{\mathcal{P}}_{E_G} \vert G \rangle=\vert G \rangle$ which follows from the definition of the projector $\hat{\mathcal{P}}_{E_G} = \delta( E_G - \hat{H} ) $. Compared with Eq.~(\ref{Heff_weak_drive}), the same Floquet low-energy Hamiltonian structure $\hat{\mathcal{P}}\hat{D}^{\dag}\hat{G}\hat{D}\hat{\mathcal{P}}$ appears in Eq.~(\ref{eq.absorbance}) if we set $\omega=\omL$ in $\alpha(\omega)$. This means that, to determine the absorbance at the driving frequency $\alpha(\omL)$, we just need to calculate the ground-state expectation value of (the RWA part of) our Floquet Hamiltonian.

When the driving frequency $\omL$ becomes large enough so that $\Delta_{{\bf q},s}$ approaches 0 (at an arbitrary momentum $\bf q$ with finite lower-band population) in Eq.~(\ref{renormalised-denominator}), the screened optical Stark shift in $\hat{\mathcal{P}}\hat{D}^{\dag}\hat{G}\hat{D}\hat{\mathcal{P}}$ will diverge (because its strength is inversely proportional to $\Delta_{{\bf q},s}$). This divergence of our Floquet Hamiltonian propagates into its ground-state expectation value, $\langle G \vert \hat{\mathcal{P}}\hat{D}^{\dag}\hat{G}\hat{D}\hat{\mathcal{P}} \vert G \rangle $. 
Thus, the absorbance spectrum $\alpha(\omL)$ in Eq.~(\ref{eq.absorbance}) will also peak (or diverge if we take $\gamma \to 0$). An optical absorption peak at in-gap frequency indicates the exciton resonance.
Consequently, the excitonic resonance frequency $\omega_{\text{ex}}$ is the smallest $\omL$ for which the screened denominator $\Delta_{{\bf q},s}$ equals to 0.

\end{document}